\documentclass{IEEEtran}
\usepackage{cite}
\usepackage{amsmath,amssymb,amsfonts}
\usepackage{algorithmic}
\usepackage{graphicx}
\usepackage{textcomp}
\usepackage{multirow}
\usepackage{makecell}
\usepackage{array}
\usepackage{threeparttable}
\usepackage[table]{xcolor}                                          
\def\BibTeX{{\rm B\kern-.05em{\sc i\kern-.025em b}\kern-.08em
		T\kern-.1667em\lower.7ex\hbox{E}\kern-.125emX}}
\begin{document}
	\title{A Conformal Transmit–Reflect Array Enabling Multifunctional Bidirectional Beam Switching for Emergency Communication System}
	\author{Yifei Liu, \IEEEmembership{Student Member, IEEE}, Fan Qin, \IEEEmembership{Member, IEEE}, Chao Gu, \IEEEmembership{Member, IEEE},
   
    Wenchi Cheng, \IEEEmembership{Senior Member, IEEE}, and Steven Gao, \IEEEmembership{Fellow, IEEE}
		\thanks{Manuscript received 17 April, 2026; revised 16 October, 2026; This work was supported by the National Key Research and Development Program under Grant of China 2024YFC3016004. (\textit{Corresponding author: Fan Qin.}) }
		\thanks{Yifei Liu, Fan Qin, and Wenchi Cheng are with the School of Telecommunications Engineering, Xidian University, Xi'an 710071, China (e-mail:louisliu@stu.xidian.edu.cn; fqin@xidian.edu.cn; wccheng@xidian.edu.cn).}
		\thanks{Chao Gu is with the School of Electronics, the Centre for Wireless Innovation, Electrical Engineering and Computer Science,  Queen’s University Belfast, U.K. (e-mail: chao.gu@qub.ac.uk).}
		\thanks{Steven Gao is with the Department of Electronic Engineering, Chinese University of Hong Kong, Hong Kong (e-mail: scgao@ee.cuhk.edu.hk).}
        \thanks{Color versions of one or more of the figures in this article are available online at http://ieeexplore.ieee.org.}}
		\maketitle
	
	\begin{abstract}
		This article presents a conformal transmit–reflect array (CTRA) that enables bidirectional multibeam switching through a single metasurface aperture for emergency communication applications. To realize reconfigurable bidirectional beam switching, a transmit–reflect integrated element (TRIE) is developed by combining a polarization-selective transmissive structure with a complementary-symmetric reflective structure. The proposed element enables independent control of the cross-polarized transmission and reflection phases while exhibiting stable performance under oblique incidence. To support practical conformal implementations, a geometrical-optics-based phase compensation method is applied to cylindrical geometries, enabling effective wavefront control across the conformal aperture. By incorporating a polarization-switchable feed array (PSFA), multifunctional beam switching is achieved in the forward, backward, and bidirectional operating modes. The PSFA is configured as a nonuniform curved linear array near the focal region, enabling feed-selective beam switching within $\pm 30^\circ$ in both the $+z$ and $-z$ directions through selective excitation of the corresponding feed elements. The full-wave simulation results agree well with the measurements. The proposed CTRA achieves peak gains of 24.1 and 24.8 dBi in the forward and backward modes, respectively. The maximum aperture efficiencies in the forward and backward operating states are 28.4$\%$ and 33.4$\%$. These results demonstrate the effectiveness and flexibility of the proposed CTRA, making it a promising candidate for adaptive emergency communication systems.

	\end{abstract}
	
	
	\section{Introduction}
	\label{sec:introduction}
	\IEEEPARstart{W}{ith} the rapid development of next-generation wireless communication systems, the demand for rapidly deployable systems that provide flexible wireless coverage continues to grow, particularly in emergency scenarios [1]–[2]. In disaster-stricken areas, communication infrastructures, such as terrestrial base stations, may be severely damaged, leading to network interruptions and coverage loss. High-altitude emergency relay platforms, including tethered balloons and emergency airships, offer an effective solution for restoring temporary communication links over large areas [3]–[5]. As illustrated in Fig. 1, conformal bidirectional apertures mounted on these platforms can function as relay nodes to establish high-gain communication links with backhaul nodes and ground rescue units. These application scenarios impose demanding requirements on antenna systems, including a low profile, lightweight configuration, low power consumption, conformal adaptability, and multifunctional beam-switching capability in both forward and backward directions [6]–[9]. Conventional beam-scanning antennas, such as lens antennas and phased arrays, can provide beam-switching capability over multiple angles; however, they generally suffer from bulky structures, complicated feeding networks, high power consumption, and deployment difficulties, which restrict their applicability to temporary emergency platforms [10]–[13]. Therefore, a highly integrated, low-power, and flexible bidirectional beam-switching antenna is highly desirable for emergency relay communications.
	
	\begin{figure}[!t]
		\centerline{\includegraphics[width=\columnwidth]{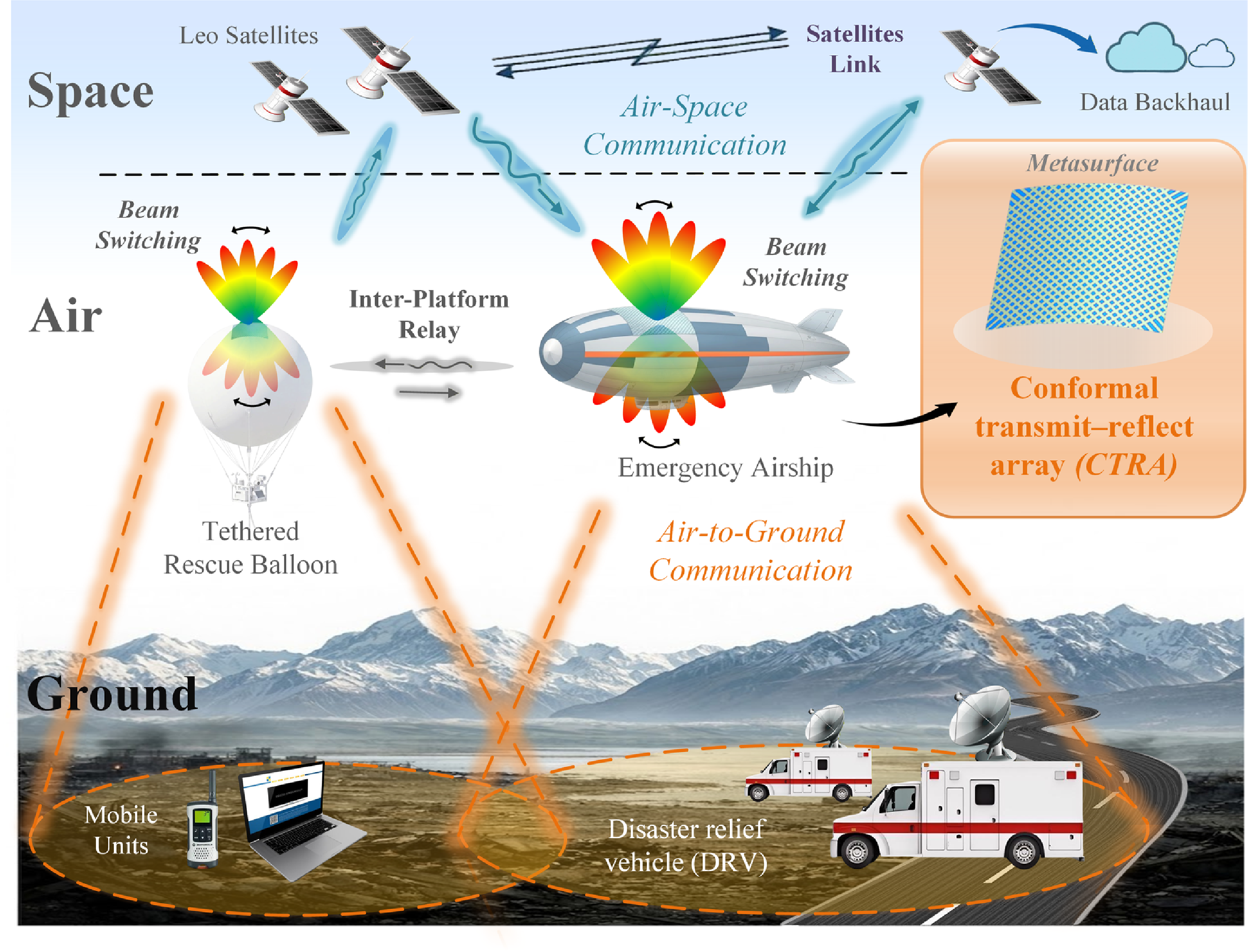}}
		\caption{Schematic representation of the application of bidirectional conformal transmit–reflect array (CTRA) within an emergency rescue network.}
		\label{fig1}
	\end{figure}
	
    Benefiting from low loss, relatively simple aperture configurations, and flexible phase-control capability, transmitarray and reflectarray antennas have been widely investigated for high-gain beam generation and beam-switching applications [14]–[18]. For example, a dual-band dual-circularly polarized transmitarray capable of precise multibeam control was reported in [18]. Nevertheless, conventional transmitarray and reflectarray antennas generally operate in a single radiation direction, which restricts their coverage capability in scenarios requiring simultaneous support for backhaul and access links. To address this limitation, bidirectional transmit–reflect arrays (TRAs) have been proposed to enhance radiation control on both sides of the aperture [19]–[28]. In our previous studies [19], [24], rapid bidirectional pencil-beam switching was realized using a hybrid transmitarray based on polarization-rotating metasurface elements and polarization grids. In addition, a metal-only reflect-transmit-array with polarization-dependent operation was reported in [20]. However, its relatively large element period, thickness, and rigid metal/air configuration limit compact integration and conformal deployment. Therefore, the development of a low-profile, multifunctional, and flexibly deployable bidirectional beam-switching metasurface suitable for relay platforms remains a critical challenge.

    Beyond bidirectional beam switching, integration with curved airborne relay platforms requires conformal apertures, motivating the transition from conventional planar transmitarrays and reflectarrays to conformal designs. Conformal metasurface antennas offer a promising solution by enabling curved-surface integration while maintaining electromagnetic performance [29]–[36]. For example, an ultrathin conformal transmitarray with segmented phase compensation was proposed in [30] for efficient beam focusing under oblique incidence. In addition, several conformal reflectarray antennas with dedicated phase compensation methods have been reported in [34]–[36], demonstrating effective reflective beam control on curved apertures. Nevertheless, most existing conformal designs rely on rigid or fixed structures and mainly support unidirectional radiation or beam switching. Therefore, a flexible, low-profile, and multifunctional conformal metasurface is highly desirable.
	
    A conformal TRA for UAV applications was recently reported in [38]; however, separate implementations are required for different radiation modes, limiting aperture integration and switching flexibility. This paper proposes a flexible conformal transmit–reflect array (CTRA) for high-altitude emergency communication relays, enabling multifunctional bidirectional multibeam operation within a single aperture. This architecture is well suited to satellite/backhaul, inter-platform, and air-to-ground emergency links, while reducing antenna duplication and associated payload burden. The present work focuses on the aperture-level design and experimental validation of conformal phase compensation, bidirectional beam switching, and multibeam radiation. Compared with existing conformal metasurfaces and beam-switching TRAs, the proposed design offers the following advantages:

	\begin{figure}[!t]
		\centerline{\includegraphics[width=\columnwidth]{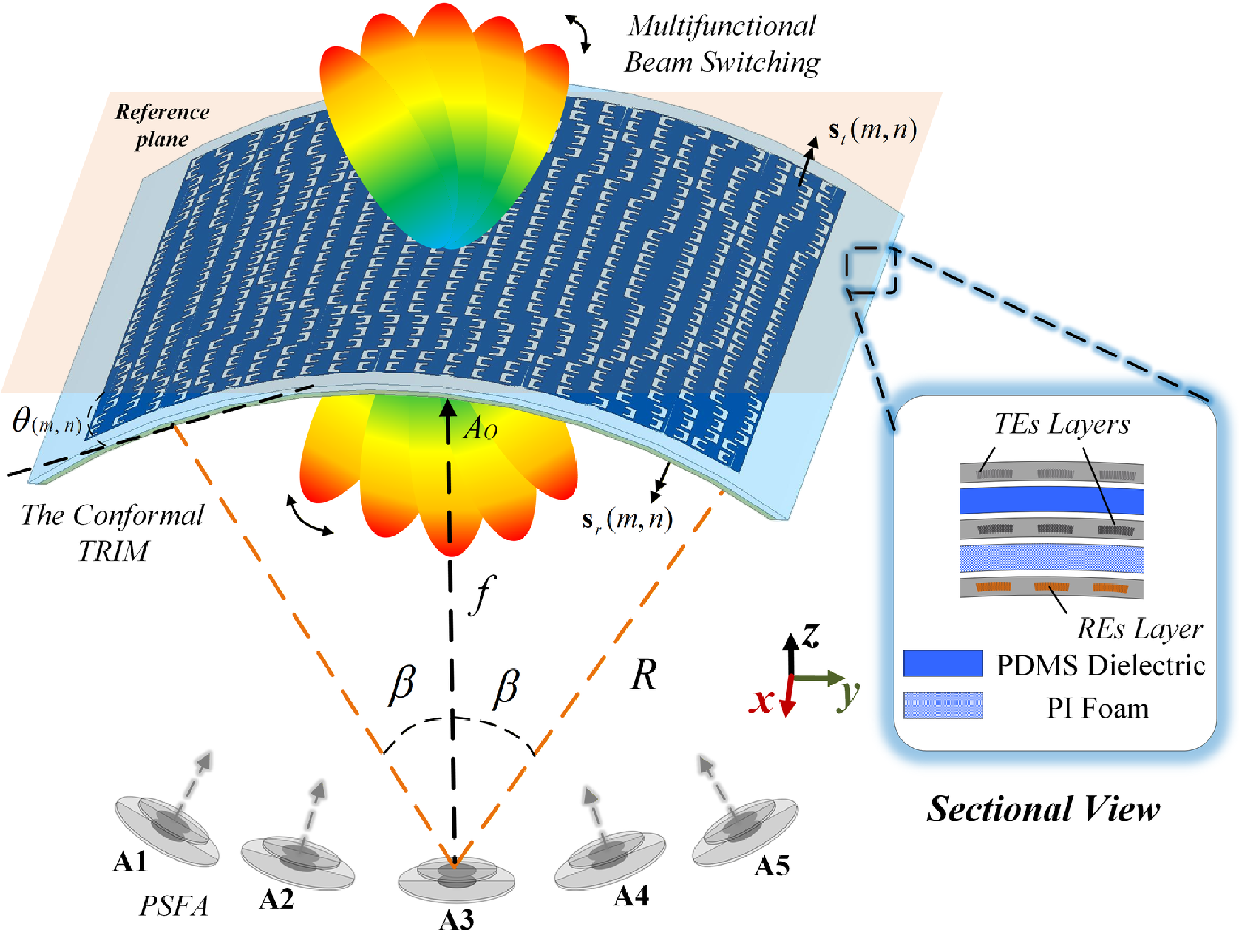}}
		\caption{Schematic configuration of the proposed CTRA and its feed arrangement.}
		\label{fig2}
	\end{figure}
	
	\begin{figure*}[!t]
		\centerline{\includegraphics[width=0.9\textwidth]{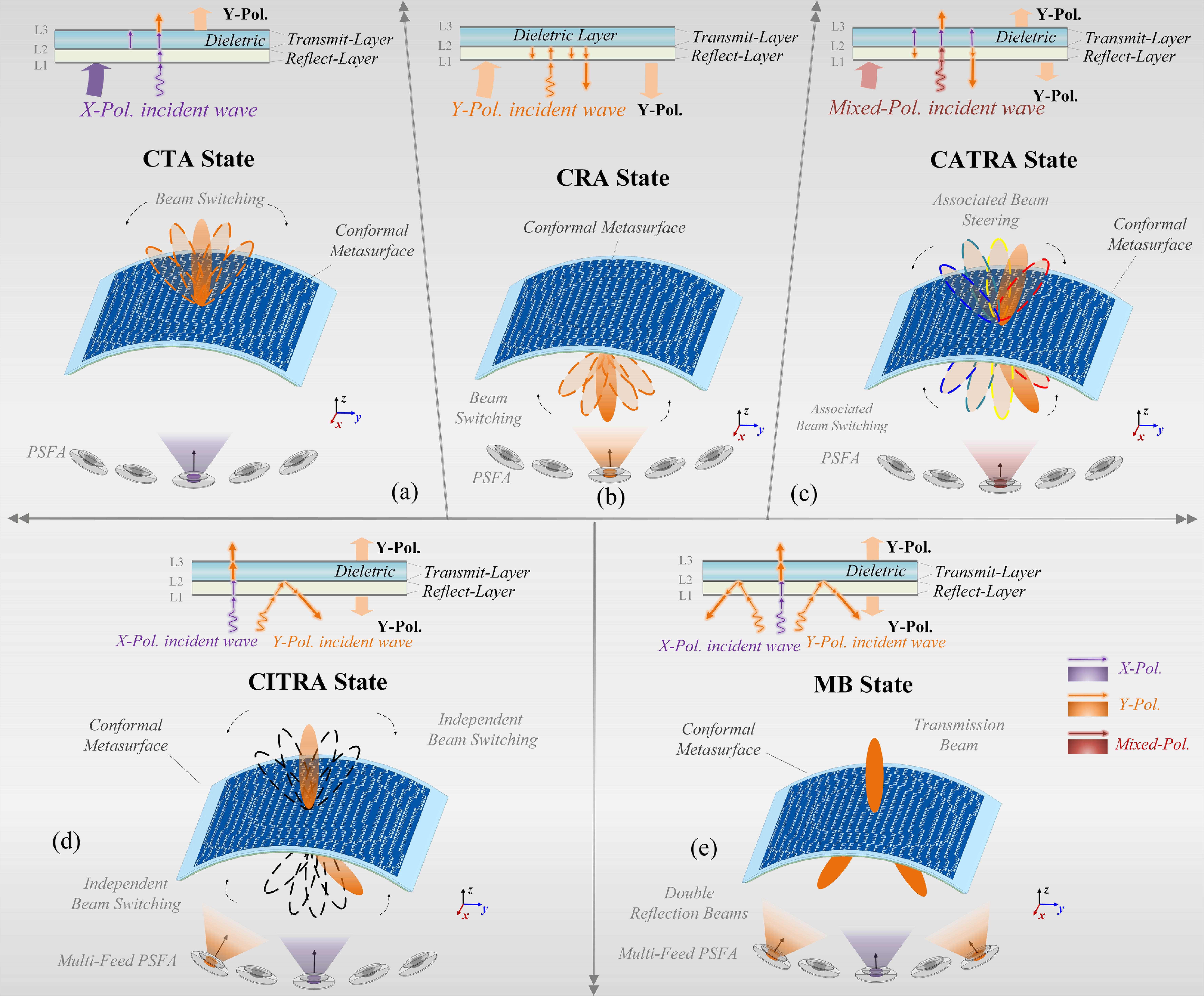}}
		\caption{The schematic diagram illustrating the principle of the proposed antenna in different working states. (a) CTA State. (b) CRA state. (c) CATRA state. (d) CITRA state. (e) MB state.}
		\label{fig3}
	\end{figure*}

    1) An ultrathin, flexible Huygens-inspired metasurface element is developed, which enables independent cross-polarized transmission and reflection phase control while exhibiting stable behavior under oblique incidences.
    
    2) A geometry-assisted phase compensation approach is developed for cylindrical conformal apertures, enabling highly integrated CTRA implementation with reduced design complexity. Combined with polarization-controlled excitation, the proposed design achieves low-sidelobe pencil-beam switching in forward, backward, and bidirectional radiation states.
    
    3) By integrating a curved polarization-switchable feed array (PSFA), the CTRA prototype enables independent bidirectional multibeam control at identical or distinct angles, with ±30° switching in both the $+z$ and $-z$ directions and simultaneous multibeam radiation.
	
	The remainder of this paper is organized as follows. Section II presents the configuration and operating principle of the proposed CTRA. Section III details the TRIE metasurface units, the design of the PSFA, and the phase compensation methodology of the CTRA. Section IV describes prototype fabrication, experimental validation, and comparative analysis with existing work. Finally, Section V concludes the paper.
	
	\section{Working Principle of the Multibeam Conformal Transmit-Reflect Array}
	
	Fig. 2 illustrates the schematic model of the proposed bidirectional multi-beam switching CTRA, which comprises a conformal transmission–reflection integrated metastructure (TRIM) conformally attached to the cylindrical surface, along with a PSFA arranged along a nonuniform arc. The TRIM integrates transmission elements (TEs) and reflection elements (REs) within a three-layer metallic structure patterned on an ultrathin dielectric substrate with intermediate dielectric layers. It enables polarization-dependent transmission or reflection of incident waves. By properly designing the metallic layers, L1 provides the required reflection phase compensation, L2 serves as the polarization-selective layer and reflection ground plane, and L3 mainly controls the transmission phase, thereby enabling high-gain transmitted and reflected pencil beams with the desired polarization characteristics. The PSFA comprises five polarization-switchable feed antennas (A1–A5) symmetrically distributed along a nonuniform arc in the yOz plane, where the distance between the central feed A3 and the center of the metasurface Ao is equal to the focal length f. In particular, the middle TRIM layer L2 exhibits linear polarization selectivity, functioning as a phase-compensation layer in transmission mode and as a ground plane in reflection mode. Because of the polarization-rotation capability of the transmission unit, the forward, backward, and bidirectional radiation states maintain identical polarization characteristics.
	
	As illustrated in Fig. 3(a), under x-polarized excitation from the PSFA, L1 has a negligible influence on transmission, while L2 and L3 jointly transform the incident wave into a focused y-polarized pencil beam. This operating mode is referred to as the conformal transmitarray (CTA) state. In contrast, under y-polarized excitation (Fig. 3(b)), L2 reflects the incident wave owing to its polarization selectivity, while L1 provides the required phase compensation to form a high-gain pencil beam along the $-z$ direction. This operating mode is referred to as the conformal reflectarray (CRA) state.

    \begin{figure}[!t]
		\centerline{\includegraphics[width=\columnwidth]{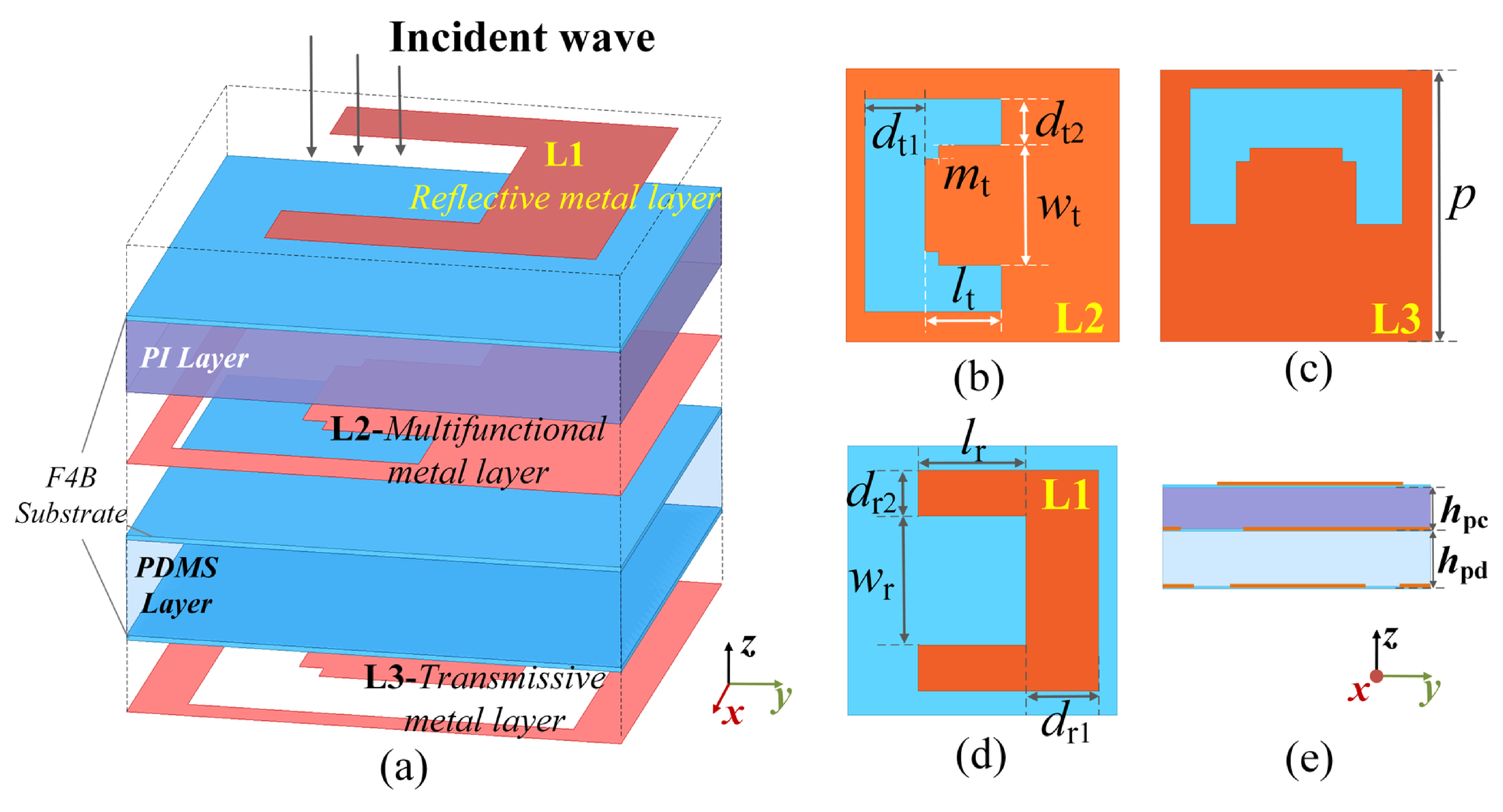}}
		\caption{ (a) The proposed TRIE element cell. (b)-(d) are transmissive/reflective metal layers. (e) Side view of the element.}
		\label{fig4}
	\end{figure}

    \begin{figure}[!t]
		\centerline{\includegraphics[width=\columnwidth]{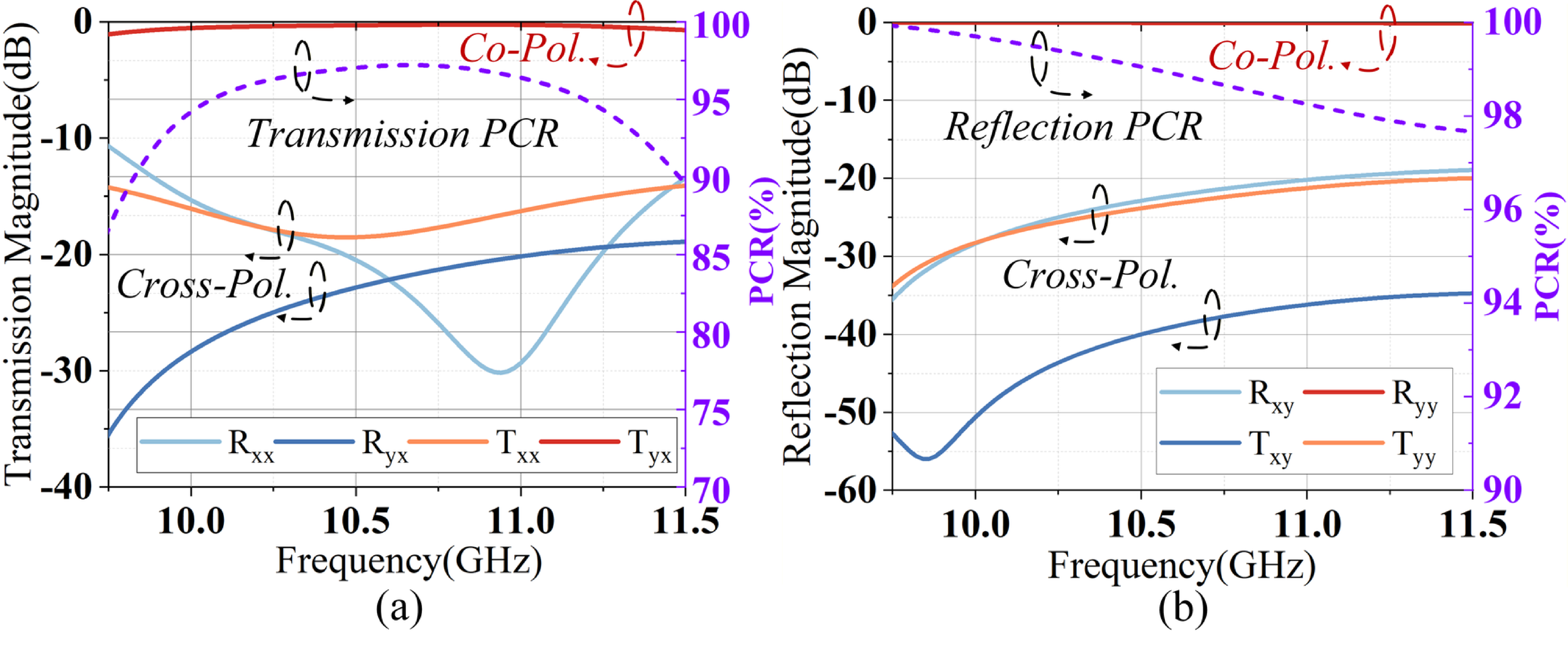}}
		\caption{ (a) Transmission, reflection coefficients, and Transmission PCR curves of TRIE when ${{\rm{l}}_{t}}$ = 5.0 mm; (b) Transmission, reflection coefficients, and Reflection PCR curves of TRIE when ${{\rm{l}}_{r}}$ = 4.0 mm.}
		\label{fig5}
	\end{figure}

    \begin{figure}[!t]
		\centerline{\includegraphics[width=\columnwidth]{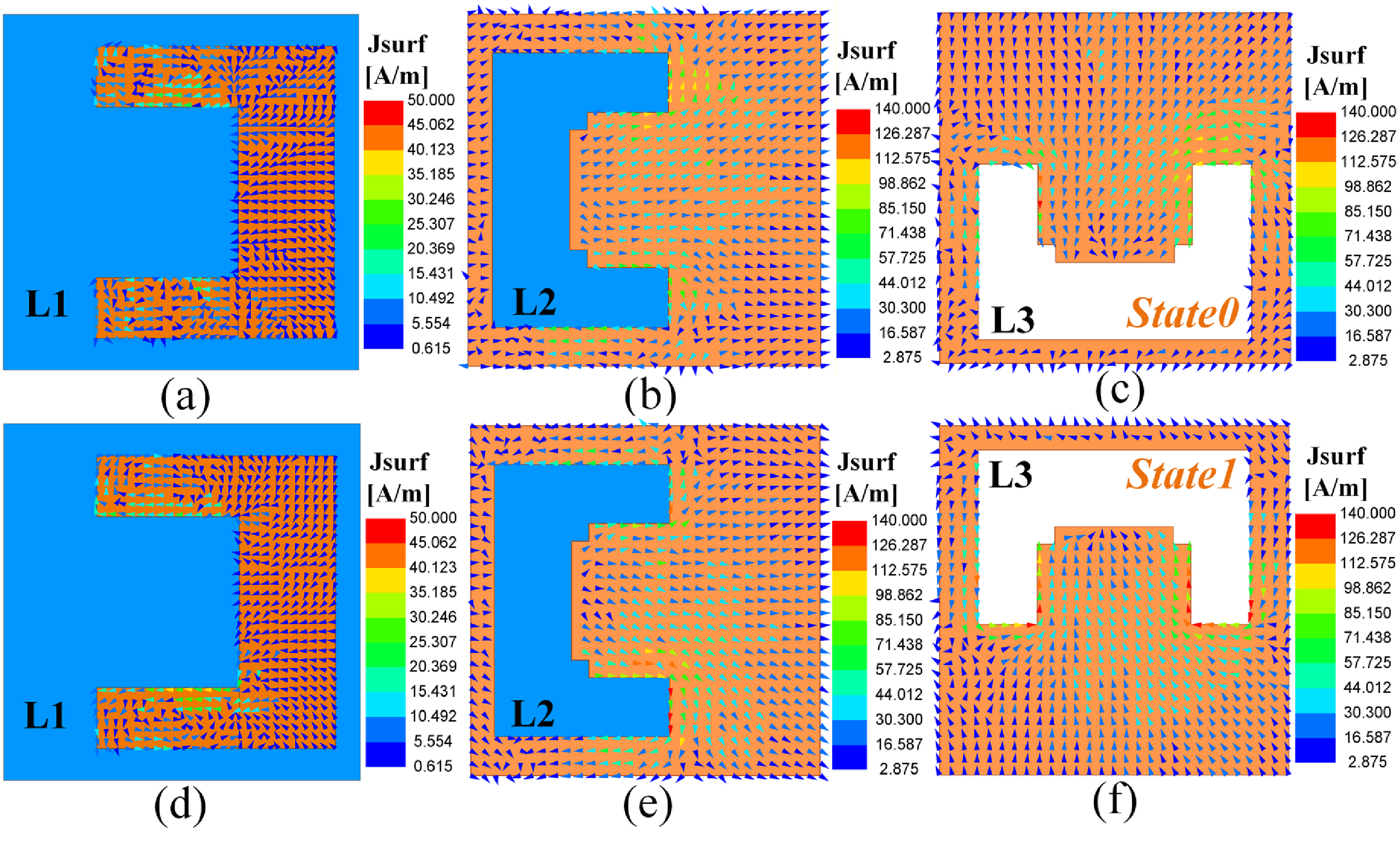}}
		\caption{ The surface-current distributions of the TRIEs. (a)-(c) State 0. (d)-(f) State 1.}
		\label{fig6}
	\end{figure}
    
	The conformal associated transmit–reflect array (CATRA) state is a superposition of the two modes: under mixed-polarization excitation, a portion of the incident energy is transmitted along $+$z, while the remainder is reflected along $-$z, producing symmetric pencil beams with the same deflection angle in both directions (Fig. 3(c)). In the conformal independent transmission–reflection array (CITRA) state, two independent feeders in the PSFA excite x- and y-polarized waves separately, enabling transmitted and reflected beams to be switched to distinct angles (Fig. 3(d)). For all four states, flexible multi-angle beam switching is achieved by selecting different PSFA excitation positions. Simultaneous x/y-polarized excitation of three feeders yields a multi-beam (MB) state with concurrent radiation of three beams (Fig. 3(e)). In Fig. 3, solid and dashed lines denote the active excitation beam and switchable beam, respectively.
	
	\section{Design of the Multibeam CTRA}
	
	\subsection{TRIE Design for the CTRA}
    As shown in Figs. 4(a)–(d), the proposed metasurface unit, referred to as a transmit–reflect integrated element (TRIE), consists of three metallic layers and multiple dielectric substrates. The top metallic layer (L1) provides phase compensation for the reflected wave, while allowing the vertically polarized incident wave to transmit without additional phase compensation. The intermediate metallic layer (L2) exhibits polarization selectivity; in the reflection state, it behaves as a metallic ground (GND) and reflects incident waves, whereas for vertically polarized waves, it permits transmission and, in conjunction with the bottom metallic layer (L3), constitutes a Huygens-like element to achieve transmitted-wave phase compensation. Each metallic layer is laminated on a dielectric F4B substrate of 0.1 mm thickness, with foam (PI) and PDMS films of thicknesses ${{\rm{h}}_{pc}}$ and ${{\rm{h}}_{pd}}$, respectively, inserted between them as dielectric spacers. As depicted in Figs. 4(b)-(e), the parameters of the element cell are: ${{\rm{d}}_{r1}}$ = 2.7 mm; ${{\rm{d}}_{r2}}$ = 1.7 mm; ${{\rm{w}}_{r}}$ = 4.8 mm; ${{\rm{d}}_{x1}}$ = 1.1 mm; ${{\rm{d}}_{y1}}$ = 0.6875 mm; ${{\rm{d}}_{t1}}$ = 2.2 mm; ${{\rm{w}}_{t}}$ = 4.4 mm; ${{\rm{d}}_{t2}}$ = 1.7 mm; ${{\rm{m}}_{t}}$ = 0.5 mm; ${{\rm{h}}_{pc}}$ = 1.5 mm; ${{\rm{h}}_{pd}}$ = 2.0 mm. The TRIE element has an overall thickness of 3.8 mm, which corresponds to approximately 0.134${{\rm{\lambda}}_{0}}$ at the center frequency of 10.6 GHz.

    \begin{figure}[!t]
		\centerline{\includegraphics[width=\columnwidth]{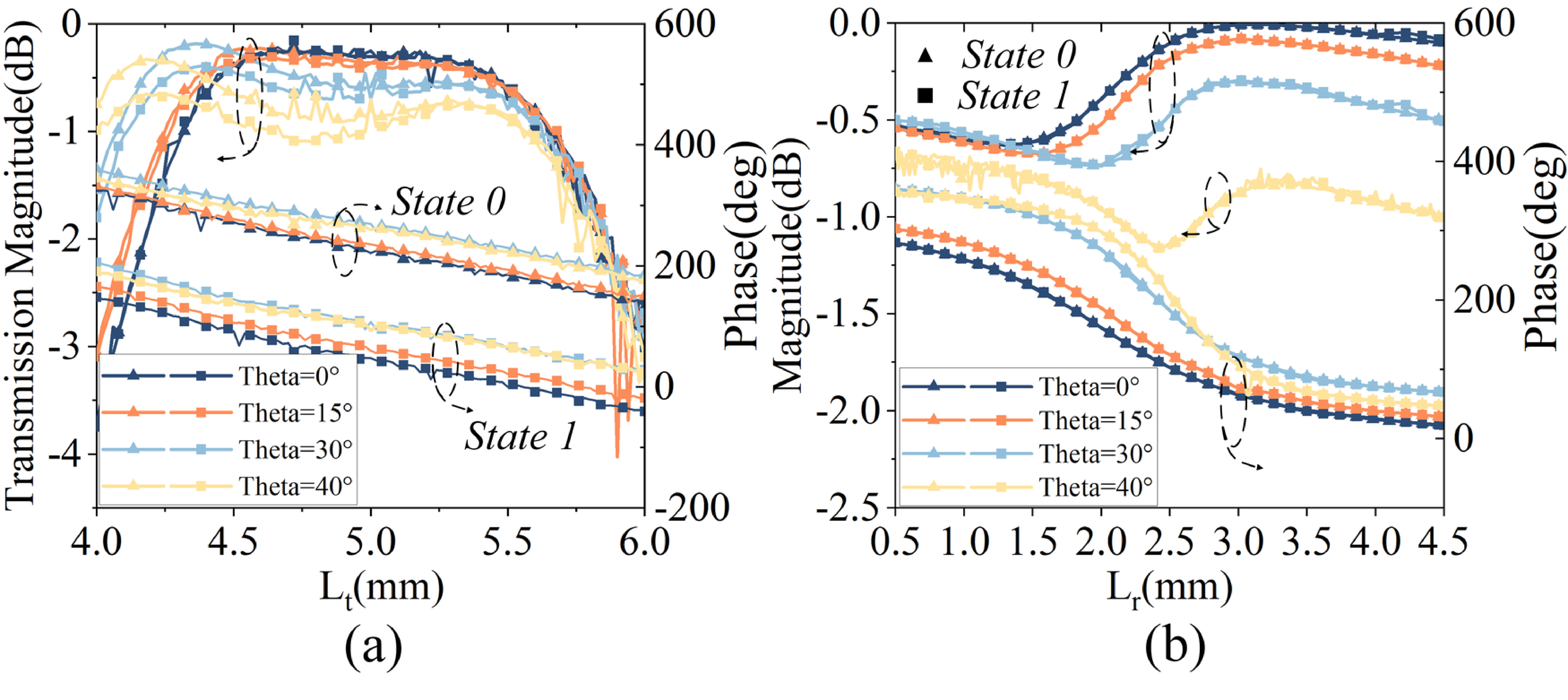}}
		\caption{ (a) Curves of transmission amplitude and phase varying with ${{\rm{l}}_{t}}$. (b) Curves of reflection amplitude and phase varying with ${{\rm{l}}_{r}}$.}
		\label{fig7}
	\end{figure}

    \begin{figure}[!t]
		\centerline{\includegraphics[width=\columnwidth]{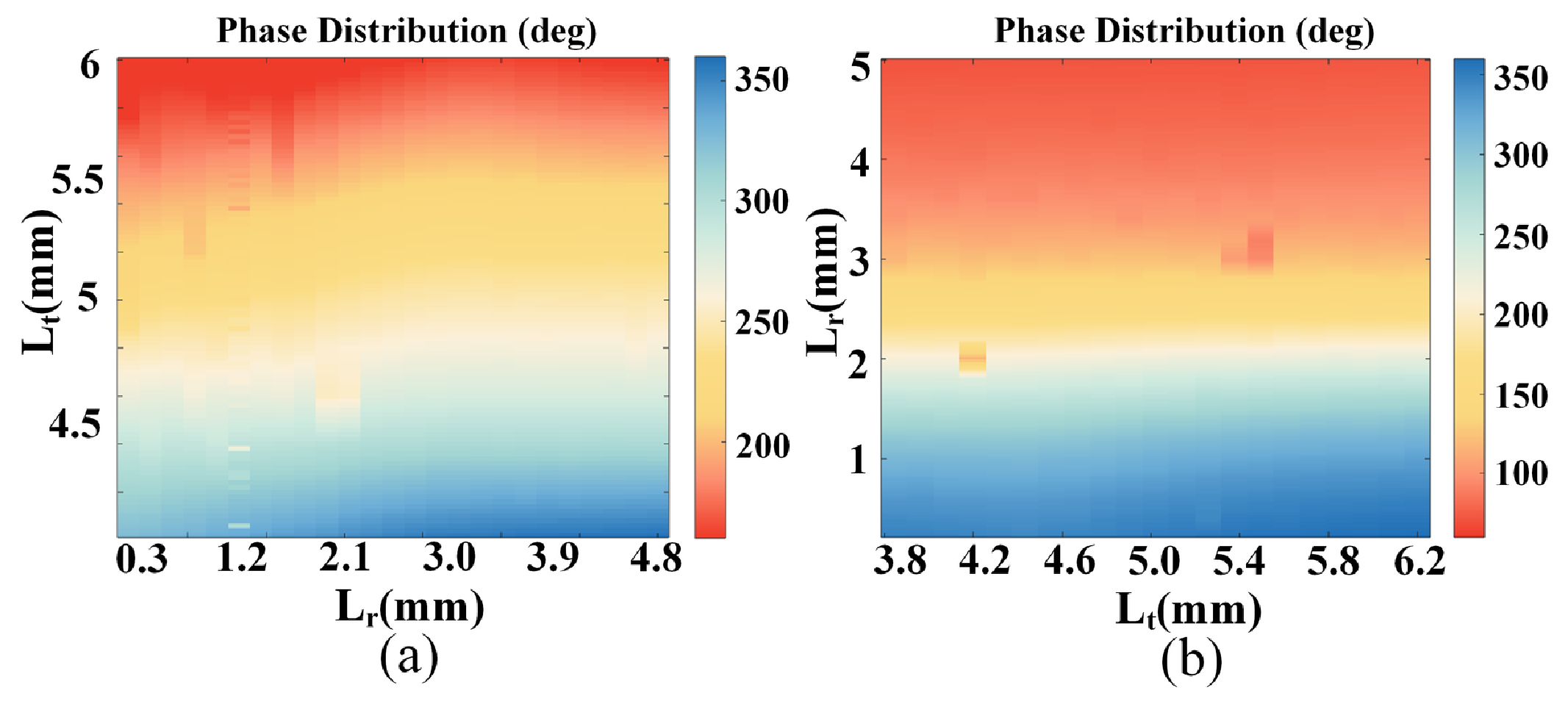}}
		\caption{ (a) Cross-sensitivity of the transmission phase to the reflection-control variable $l_r$. (b) Cross-sensitivity of the reflection phase to the transmission-control variable $l_t$.}
		\label{fig8}
	\end{figure}
    
	To evaluate the polarization selection performance of the TRIEs within the target frequency band, the unit cell was optimized in Ansys HFSS using master-slave boundaries and Floquet-port excitation. As illustrated in Fig. 5, the amplitude response of the TRIE units is presented as a function of frequency. Specifically, ${{\rm{T}}_{yx}}$ and ${{\rm{R}}_{yy}}$ denote the transmission amplitude of the x-polarized wave and the reflection amplitude of the y-polarized wave, respectively. Within the frequency range of 9.75–11.50 GHz, ${{\rm{T}}_{yx}}$ and ${{\rm{R}}_{yy}}$ are both stable and maintain magnitudes close to 0 dB. The remaining parameters ${{\rm{T}}_{xx}}$, ${{\rm{T}}_{xy}}$, and ${{\rm{T}}_{yy}}$ are defined as transmission coefficients, while ${{\rm{R}}_{xx}}$ and ${{\rm{R}}_{yx}}$ are defined as reflection coefficients. The polarization conversion rates (PCRs) for transmission and reflection are defined as follows:
    \begin{equation}
	{PCR_{(t)}} = {\left| {{T_{yx}}} \right|^2}/({\left| {{T_{yx}}} \right|^2} + {\left| {{T_{xx}}} \right|^2} + {\left| {{R_{yx}}} \right|^2} + {\left| {{R_{xx}}} \right|^2})
    \end{equation}
    \begin{equation}
	{PCR_{(r)}} = {\left| {{R_{yy}}} \right|^2}/({\left| {{R_{yy}}} \right|^2} + {\left| {{R_{xy}}} \right|^2} + {\left| {{T_{xy}}} \right|^2} + {\left| {{T_{yy}}} \right|^2})
    \end{equation}

     \begin{figure}[!t]
		\centerline{\includegraphics[width=\columnwidth]{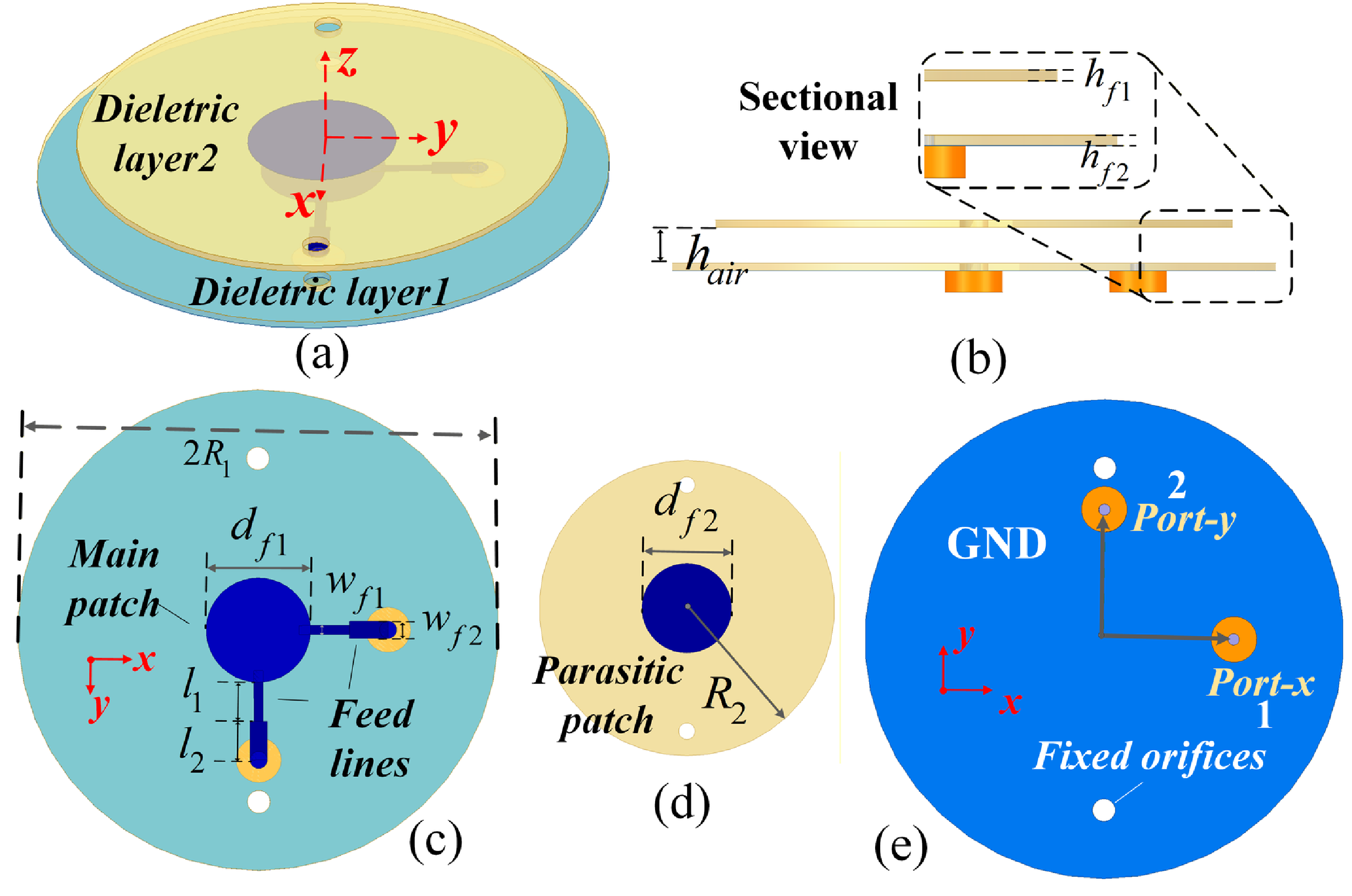}}
		\caption{ (a) Configuration of the feed antenna. (b) Side view of the feed antenna. (c) The feed network and radiation patch structure of the antenna. (d) Parasitic patch structure of the antenna. (e) Different polarization excitation ports of the feed. The parameters of the feed antenna are: ${{\rm{R}}_{1}}$ = 21 mm; ${{\rm{R}}_{2}}$ = 18 mm; ${{\rm{d}}_{f1}}$ = 9.12 mm; ${{\rm{d}}_{f2}}$ = 10.92 mm; ${{\rm{w}}_{f1}}$ = 0.8 mm; ${{\rm{w}}_{f2}}$ = 1.5 mm; ${{\rm{l}}_{1}}$ = 3.32 mm; ${{\rm{l}}_{2}}$ = 3.34 mm; ${{\rm{h}}_{f1}}$ = ${{\rm{h}}_{f2}}$ = 0.5 mm; ${{\rm{h}}_{air}}$ = 2.5 mm.}
		\label{fig9}
	\end{figure}
    
    \begin{figure}[!t]
		\centerline{\includegraphics[width=\columnwidth]{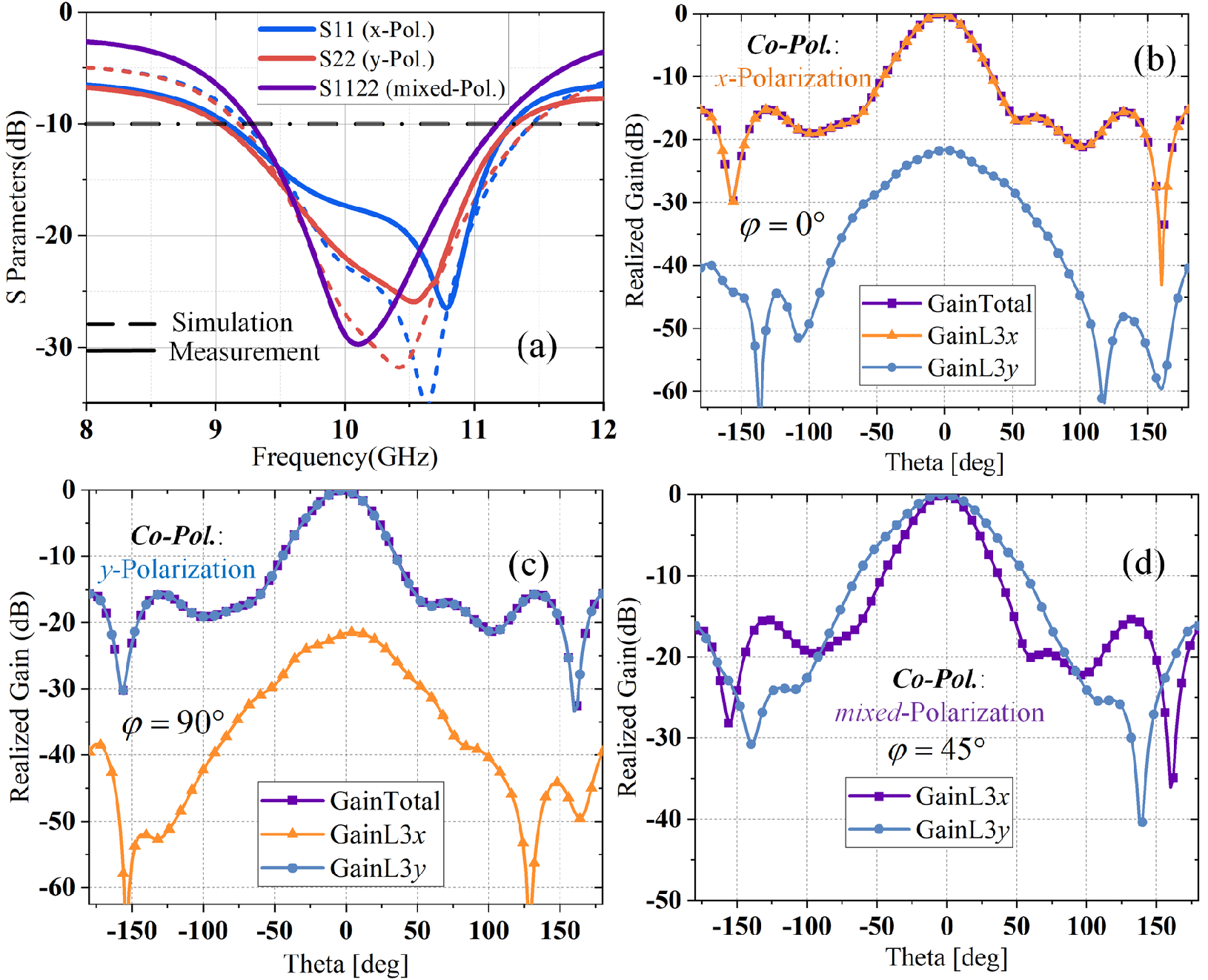}}
		\caption{ (a) Simulated and measured S-parameters of the PSFA in different states. (b) Radiation patterns of the PSFA in x-polarization. (c) Radiation patterns of the PSFA in y-polarization. (d) Radiation patterns of the PSFA in mixed polarization.}
		\label{fig10}
	\end{figure}   
    
 	The curve of ${PCR_{(t)}}$ in the transmission state is shown in Fig. 5(a). Within the frequency range of 9.8-11.5 GHz, the value of ${PCR_{(t)}}$ remains above 90$\%$. At the same time, the value of ${PCR_{(r)}}$ is always no less than 97$\%$ within the frequency band of 9.75 - 11.5 GHz, as shown in Fig. 5(b). The transmission and reflection PCRs differ because the two channels employ different dielectric stacks and complementary-symmetric metallic patterns; nevertheless, both remain above their specified thresholds over the operating band. As shown in Fig. 6, the electric field distributions of the TRIEs are presented, where a 180° mirror symmetry on the L3 surface is observed between states 0 and 1. As shown in Fig. 7(a), the transmission amplitude and phase curves of TRIE with respect to the length ${{\rm{l}}_{t}}$ vary from 4.5 to 5.5 mm, and a phase coverage of 180 ° can be achieved. Due to the symmetrical structure of the unit itself, by rotating L3 180°, another 180° phase coverage can be obtained. Similarly, Fig. 7(b) shows the curves of the reflection amplitude and phase with respect to ${{\rm{l}}_{r}}$. As ${{\rm{l}}_{r}}$ varies from 0.5 to 4.0 mm, a 360° phase coverage can be achieved. 
    
 Due to the unique structural configuration, the transmission and reflection phases are primarily controlled by $l_t$ and $l_r$, respectively. To quantitatively evaluate the mutual coupling between the two phase-control parameters, a cross-sensitivity analysis is performed based on the phase perturbations obtained from the two-dimensional parameter sweep in Fig. 8. Before calculating the phase variations, the simulated phase responses are unwrapped to eliminate the artificial discontinuity at $0^{\circ}/360^{\circ}$. The maximum cross-induced phase perturbations are defined as

\begin{equation}
\Delta \phi_{t,\max}
=
\max_{l_t}
\left[
\max_{l_r}\phi_t(l_t,l_r)
-
\min_{l_r}\phi_t(l_t,l_r)
\right],
\end{equation}

\begin{equation}
\Delta \phi_{r,\max}
=
\max_{l_r}
\left[
\max_{l_t}\phi_r(l_t,l_r)
-
\min_{l_t}\phi_r(l_t,l_r)
\right],
\end{equation}

where $\phi_t(l_t,l_r)$ and $\phi_r(l_t,l_r)$ denote the unwrapped transmission and reflection phases, respectively. $\Delta \phi_{t,\max}$ represents the maximum perturbation of the transmission phase caused by varying the reflection-control parameter $l_r$, whereas $\Delta \phi_{r,\max}$ represents the maximum perturbation of the reflection phase induced by varying the transmission-control parameter $l_t$. Smaller values of these two quantities indicate weaker cross-coupling and, consequently, better independent phase-control capability.

As shown in Fig. 8, the transmission phase remains nearly unchanged as $l_r$ varies from 0.2 to 5.0 mm, while the reflection phase exhibits only slight variation as $l_t$ varies from 4.0 to 6.0 mm. These results indicate weak mutual coupling between the transmission and reflection responses, allowing $l_t$ and $l_r$ to serve as approximately independent control parameters for beam synthesis and optimization. The proposed complementary-symmetric quasi-Huygens TRIE integrates independent transmission and reflection phase control into a compact, low-profile, and flexible unit cell, with its reduced period enabling finer phase sampling for conformal wavefront synthesis.
    
	\subsection{Feed Antenna Design}
	The operating states of the CTRA are enabled by a polarization-switchable feed array (PSFA). As illustrated in Figs. 9(a) and 9(b), each PSFA element comprises two dielectric substrates, a metallic feeding network, and two metallic patch layers. The lower driven patch and feeding network are shown in Fig. 9(c), whereas the upper parasitic patch is presented in Fig. 9(d). Fig. 9(e) depicts the bottom feeding ports and metallic ground plane. When Port 1 and Port 2 are excited individually, the PSFA operates in the x‑polarization mode and the y‑polarization mode, respectively. When both ports are simultaneously excited through a power divider with equal amplitudes and phases, the PSFA element generates a linearly polarized field oriented at 45°, thereby providing a mixed-polarization excitation state.
    
	Fig. 10(a) illustrates the variation of the S-parameters of the PSFA under different polarization states. The S-parameters remains below –10 dB in the frequency range of 9.3 to 11.2 GHz. Figs. 10(b)–(d) present the radiation patterns of the main polarization and cross-polarization components in the x-polarization, y-polarization, and mixed-polarization states, respectively.  In the x-polarization and y-polarization states, the cross-polarization level remains below -20 dB.
	
	\subsection{Phase Distribution of CTA, CRA, and CTRA}
	
	Phase synthesis for a conformal aperture is more complicated than that for a planar aperture because the feed-to-element distance and local surface orientation vary across the curved surface. Existing phase compensation methods for conformal transmitarrays/reflectarrays can generally be classified into two categories. The first method divides the conformal aperture into several angular regions and applies different phase corrections to each region [30]. Although computationally simple, this piecewise treatment may introduce considerable phase discretization errors. The second method first determines the phase distribution on a tangent projection plane and then introduces additional corrections associated with the curved geometry and oblique propagation [38]. This method provides element-level phase compensation but requires multiple calculation steps. In this work, a simplified geometrical-optics-based projection method is adopted to obtain the conformal phase distribution with reduced computational complexity. Its applicability is further evaluated through the residual phase error and full-wave radiation results.
        
    \begin{figure}[!t]
		\centerline{\includegraphics[width=0.85\columnwidth]{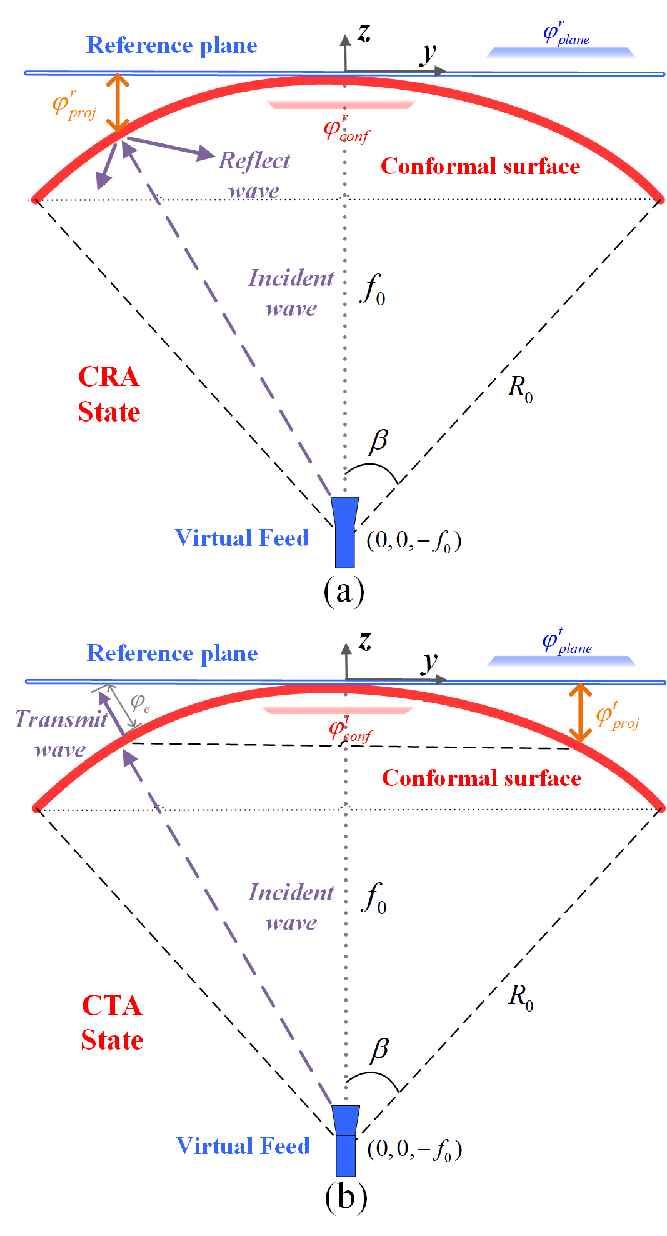}}
		\caption{  (a) The phase compensation principle of the reflection wave under the CRA state. (b) The phase compensation principle of the transmission wave under the CTA state.}
		\label{fig11}
	\end{figure}  
	
	The phase synthesis principles for the CRA and CTA states are illustrated in Fig. 11(a) and (b), respectively. The cylindrical conformal aperture has a radius $R$, and its tangent plane at the aperture center is selected as the reference projection plane.

    \begin{figure}[!t]
		\centerline{\includegraphics[width=\columnwidth]{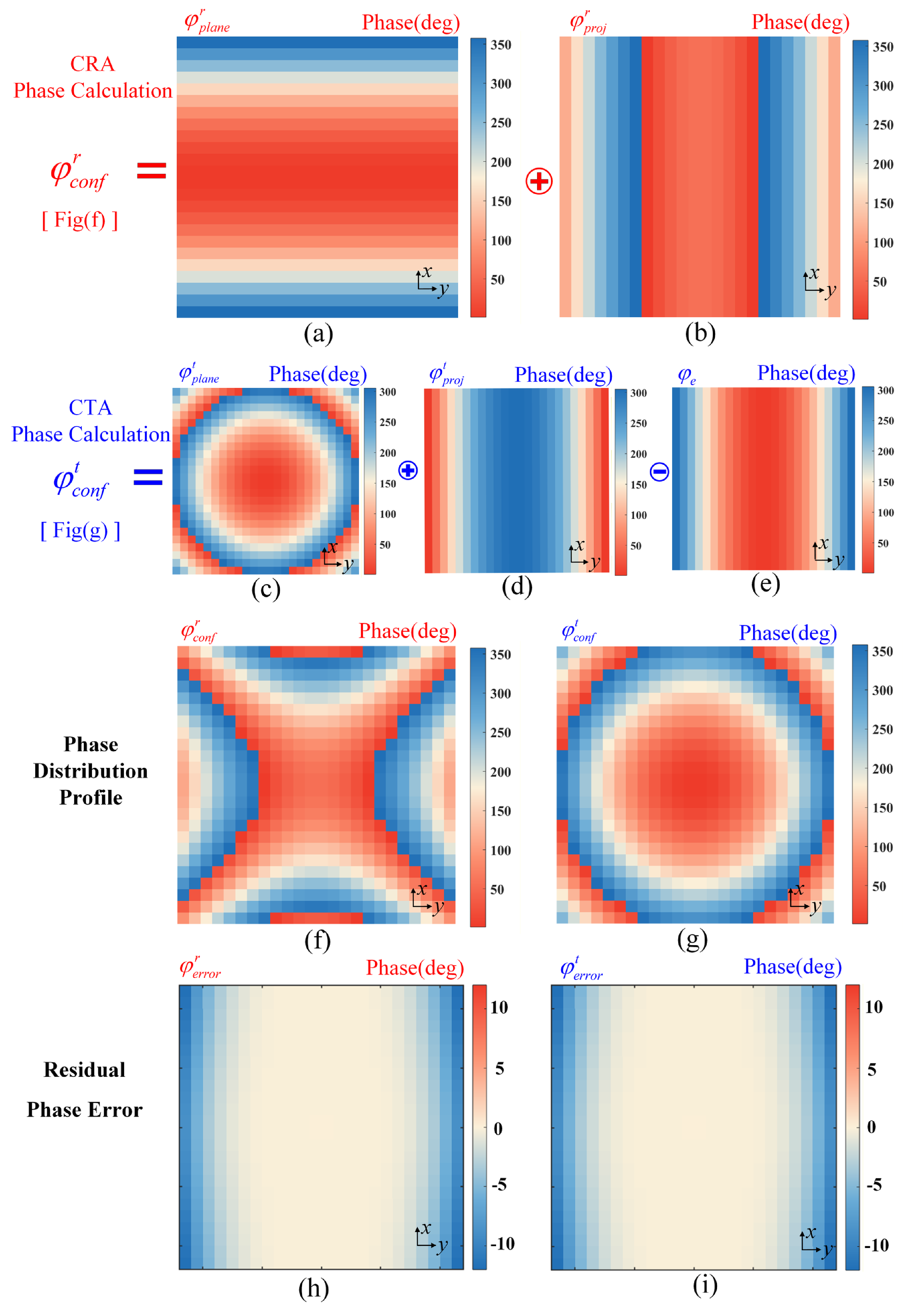}}
		\caption{  The calculation result of phase compensation (a) $\varphi _{plane}^r$. (b) $\varphi _{proj}^r$. (c) $\varphi _{plane}^t$. (d) $\varphi _{proj}^t$. (e) ${\varphi _e}$. (f) Phase distribution of CRA $\varphi _{conf}^r$. (g) Phase distribution of CTA $\varphi _{conf}^t$. (h) and (i) are residual phase error of CRA and CTA. }
		\label{fig12}
	\end{figure}  
    
For the CRA state, the virtual feed is located at a focal distance $f_0$ from the conformal aperture center. Let $s_n$ denote the signed arc length of the $n$th element row from the aperture center, with the corresponding central angle $\theta_n=s_n/R$. The local radial line through the conformal element intersects the tangent projection plane at $y_n=R\tan\theta_n$, yielding the projected coordinates $(x_m,y_n,0)$ for the $(m,n)$th element. Since the cylindrical surface is curved only along the $y$-direction, the planar reference phase is given by

\begin{equation}
\varphi_{\mathrm{plane}}^{r}(m,n)
=
\frac{2\pi}{\lambda}
\left(
\sqrt{x_m^2+y_n^2+f_0^2}
-f_0
\right)
\end{equation}

where $\lambda$ is the free-space wavelength. It should be noted that $s_n$ represents the physical arc-length coordinate on the cylindrical aperture, whereas $y_n$ represents the corresponding coordinate on the tangent projection plane.

The curvature introduces an additional path-length difference between each conformal element and its corresponding point on the projection plane. According to the above geometrical relationship, the local angular coordinate is defined as

\begin{equation}
\theta_n
=
\frac{s_n}{R}
=
\tan^{-1}\left(\frac{y_n}{R}\right)
\end{equation}

and the axial projection of the local radius is

\begin{equation}
R\cos\theta_n
=
\frac{R^2}{\sqrt{R^2+y_n^2}}
\end{equation}

Therefore, the curvature-induced phase correction in the CRA state is given by

\begin{equation}
\varphi_{\mathrm{proj}}^{r}(n)
=
\frac{2\pi}{\lambda}
\left(
\frac{R^2}{\sqrt{R^2+y_n^2}}
-R\cos\beta
\right)
\end{equation}

where $\beta=\max_n|\theta_n|$ denotes the maximum feed-illumination angle along the $y$--$z$ plane. For a cylindrical aperture with an arc length $L_y$ and a symmetric configuration, $\beta=L_y/(2R)$. The term $R\cos\beta$ is a constant reference distance and does not affect the synthesized radiation pattern.

Accordingly, the required phase distribution in the CRA state is

\begin{equation}
\varphi_{\mathrm{conf}}^{r}(m,n)
=
\operatorname{mod}
\left[
\varphi_{\mathrm{plane}}^{r}(m,n)
+
\varphi_{\mathrm{proj}}^{r}(n),
2\pi
\right]
\end{equation}
    
For the CTA state, the phase distribution $\varphi_{\mathrm{plane}}^{t}$ is first synthesized on the tangent projection plane according to the desired transmitted-beam direction. The axial path-length difference between the cylindrical aperture and the projection plane is

\begin{equation}
\Delta l_{\mathrm{proj}}^{t}(n)
=
R-R\cos\theta_n
\end{equation}

Thus, the corresponding projection correction is

\begin{equation}
\varphi_{\mathrm{proj}}^{t}(n)
=
\frac{2\pi}{\lambda}
\left(
R-\frac{R^2}{\sqrt{R^2+y_n^2}}
\right)
\end{equation}

In addition, the field transmitted by a locally tilted conformal element acquires an extra propagation phase before reaching the tangent reference plane. Along the local radial direction, the distances from the cylinder center to the tangent plane and the conformal element are $R\sec\theta_n$ and $R$, respectively. Hence, the additional propagation distance is

\begin{equation}
\Delta l_e(n)
=
R\sec\theta_n-R
=
\sqrt{R^2+y_n^2}-R
\end{equation}

The corresponding forward propagation phase is

\begin{equation}
\varphi_e(n)
=
\frac{2\pi}{\lambda}
\left(
\sqrt{R^2+y_n^2}-R
\right)
\end{equation}

Therefore, $\varphi_{\mathrm{proj}}^{t}$ represents the phase correction associated with the axial separation between the conformal aperture and the tangent projection plane, whereas $\varphi_e$ represents the additional phase accumulated by the transmitted wave along the local radial direction before reaching the tangent reference plane. The required phase distribution in the CTA state is consequently expressed as

\begin{equation}
\varphi_{\mathrm{conf}}^{t}(m,n)
=
\operatorname{mod}
\left[
\varphi_{\mathrm{plane}}^{t}(m,n)
+
\varphi_{\mathrm{proj}}^{t}(n)
-
\varphi_e(n),
2\pi
\right]
\end{equation}

  The residual phase error is evaluated as
\begin{equation}
\Delta\phi_{\mathrm{err}}^{q}(m,n)
=
\angle\left\{
\exp\left[
j\left(
\phi_{\mathrm{conf}}^{q}(m,n)
-
\phi_{\mathrm{ref}}^{q}(m,n)
\right)
\right]
\right\},
\label{eq:phase_error}
\end{equation}

     \begin{figure}[!t]
		\centerline{\includegraphics[width=\columnwidth]{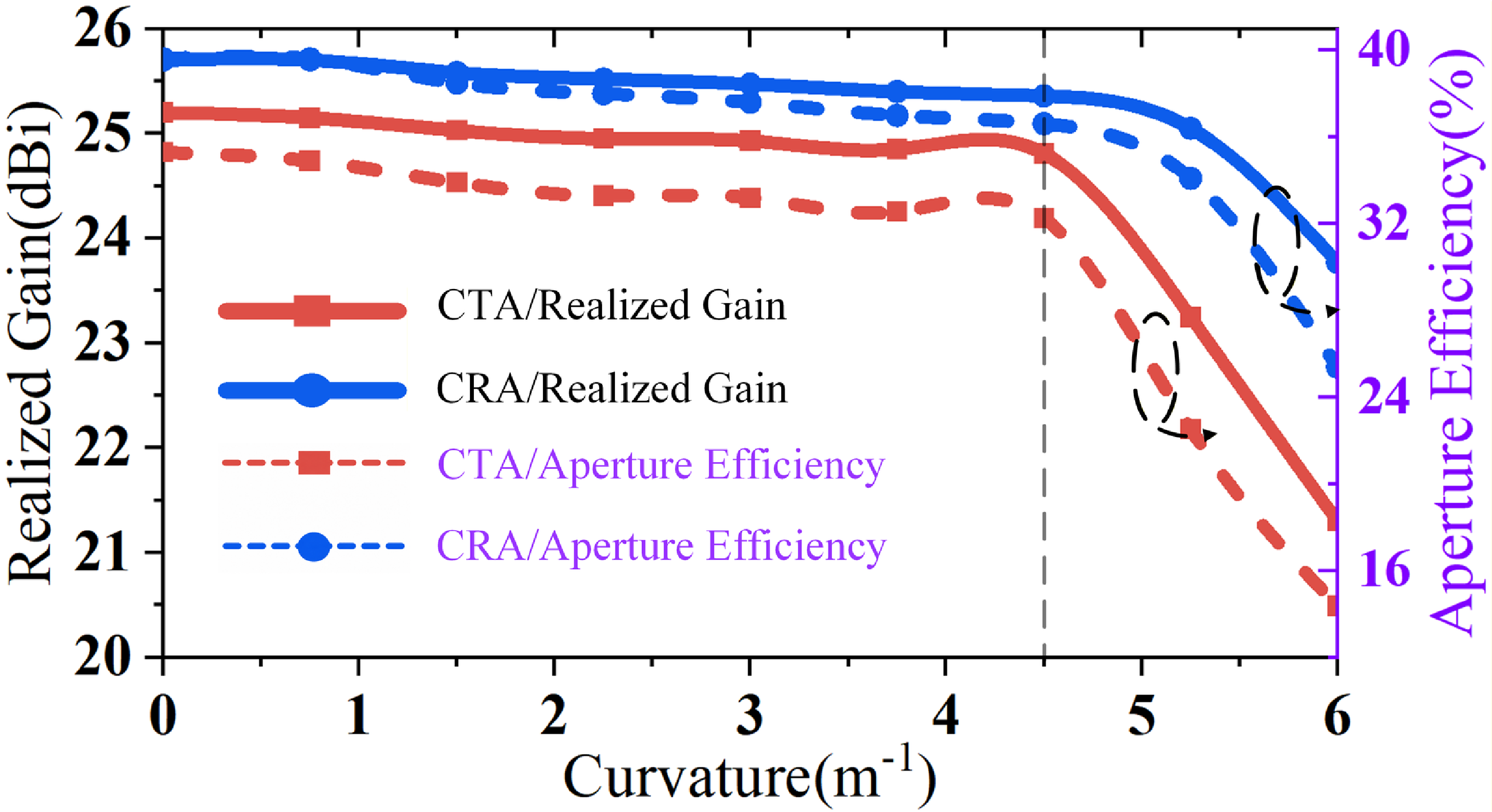}}
		\caption{ Simulated gain and aperture efficiency of the CTA and CRA states as functions of the CTRA curvature.}
		\label{fig13}
	\end{figure}

    \begin{figure}[!t]
		\centerline{\includegraphics[width=\columnwidth]{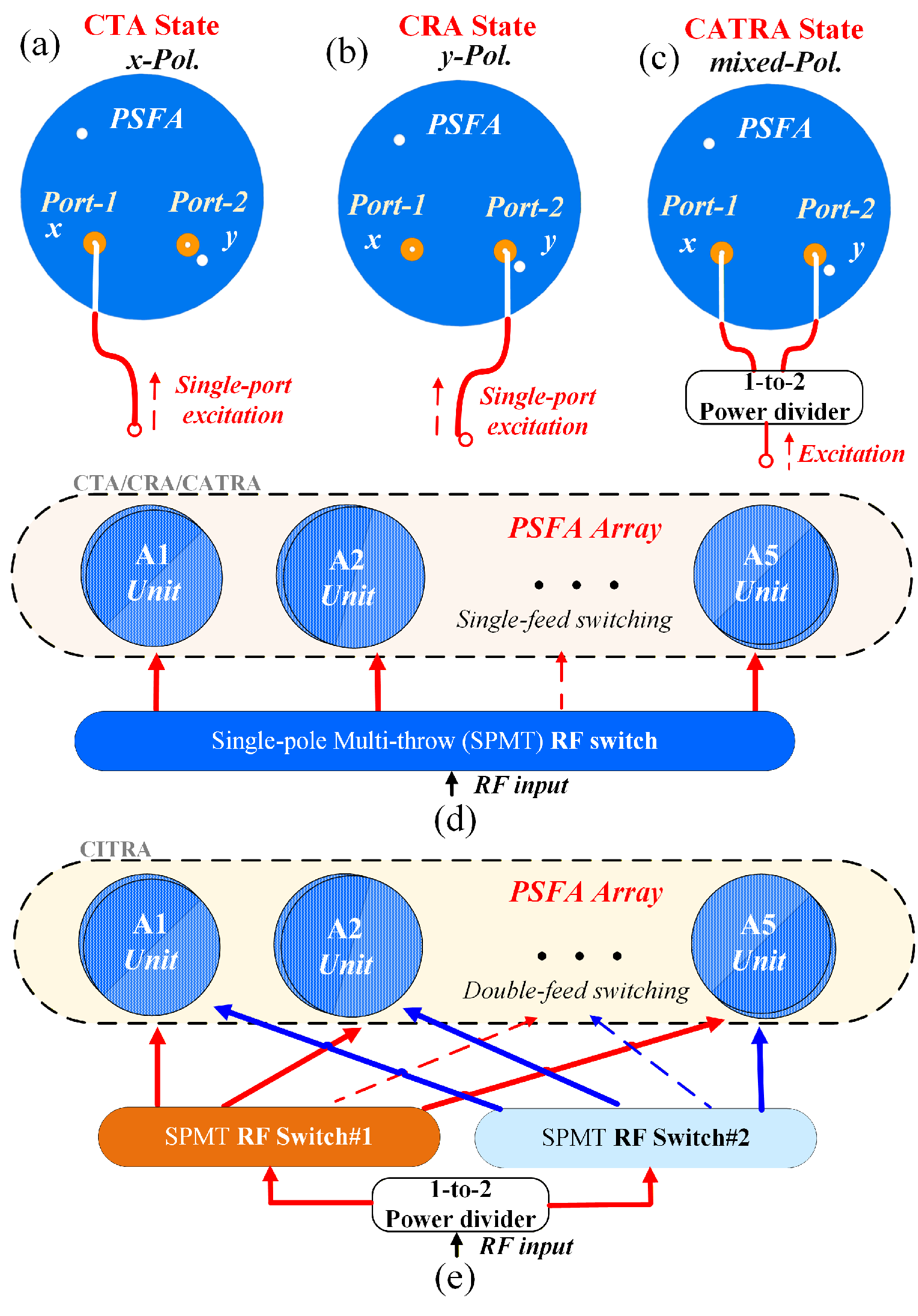}}
		\caption{(a) PSFA port connection for the CTA state. (b) PSFA port connection for the CRA state. (c) PSFA port connection for the CATRA state. (d) Beam-switching scheme for the CTA, CRA, and CATRA states. (e) Beam-switching scheme for the CITRA state.}
		\label{fig14}
	\end{figure}
    
where $q=r$ and $q=t$ denote the CRA and CTA states, respectively. The exact continuous reference phase $\phi_{\mathrm{ref}}^{q}$ is calculated from the actual cylindrical element coordinates and feed-to-element distances, with the mean optical path of the four central elements taken as the common phase reference. The operator $\angle(\cdot)$ wraps the phase difference into $[-\pi,\pi)$.

  The calculated phase components and final phase distributions for the CRA and CTA states are shown in Fig. 12. As a first-order geometrical-optics formulation, it exhibits residual errors near the aperture edges due to increased local incidence angles. The residual-error maps quantify the approximation accuracy, while the subsequent full-wave far-field results further validate the resulting radiation performance for the present curvature and switching range.

	To further assess the curvature dependence of the proposed CTRA, Fig.~13 presents the simulated gain and aperture efficiency of the CTA and CRA states at different curvatures. For each curvature, the conformal geometry and phase compensation were recalculated while keeping the aperture size, focal distance, and feed positions unchanged. The aperture efficiency was evaluated using the actual curved aperture area. Relative to the planar case ($k=0$), the gain reductions at $k=4.5~\mathrm{m}^{-1}$ are only 0.33 and 0.34 dB for the CTA and CRA states, respectively. The gain degradation remains below 1~dB up to $k=4.5~\mathrm{m}^{-1}$ for CTA and $5~\mathrm{m}^{-1}$ for CRA. Further increasing the curvature enlarges the geometric deformation and local incidence angles, thereby reducing the gain and aperture efficiency. Therefore, $k=4.5~\mathrm{m}^{-1}$ is adopted in the proposed design.

   For multiple-feed operation, the PSFA adopts a non-uniform arc-shaped arrangement to realize flexible beam switching. The off-axis feed positions are optimized according to the cylindrical aperture geometry and prescribed switching directions to reduce the residual phase error and improve the aperture phase matching.

For the $i$-th off-axis feed, the incident phase at the $(m,n)$-th element is

\begin{equation}
\phi_{m,n}^{\mathrm{inc},i}=k_0 r_{m,n}^{f_i},
\end{equation}

where $r_{m,n}^{f_i}$ is the distance from the feed to the element. For the desired switching direction $\hat{\mathbf{u}}_i$, the equivalent ideal propagation distance is defined as

\begin{equation}
r_{m,n}^{\mathrm{ref},q}
=
r_{0}^{\mathrm{ref},q}
+\hat{\mathbf{u}}_i\cdot(\mathbf{s}_{m,n}-\mathbf{s}_{0})
-\frac{\phi_{m,n}^{q}-\phi_{0}^{q}}{k_0},
\end{equation}

where $q\in\{\mathrm{CTA},\mathrm{CRA}\}$ and $\phi_{m,n}^{q}$ denotes the corresponding phase compensation of the CTRA. The wrapped feed-induced phase mismatch is calculated as

\begin{equation}
\Delta\phi_{feed}^{i,q}
=
\angle\left\{
\exp\left[jk_0
\left(r_{m,n}^{f_i}-r_{m,n}^{\mathrm{ref},q}\right)
\right]
\right\}.
\end{equation}
	
The feed position parameter $\beta_i$ is optimized by minimizing the RMS residual phase mismatch over the aperture,

\begin{equation}
\beta_i^{\mathrm{opt}}
=
\arg\min_{\beta_i\in\Omega_i}
\sqrt{
\frac{1}{MN}
\sum_{m,n}
\left|\Delta\phi_{feed}^{i,q}\right|^2
}.
\end{equation}
    
The optimization is performed at 10.6 GHz using the centers of all $24\times24$ elements with equal weighting. A one-dimensional search within the prescribed feed-position range $\Omega_i$ is employed, while each feed is oriented toward the center of the conformal aperture. The RMS and maximum residual phase mismatches are subsequently evaluated to quantify the effectiveness of the optimized feed arrangement.

\begin{figure*}[!t]
		\centerline{\includegraphics[width=0.9\textwidth]{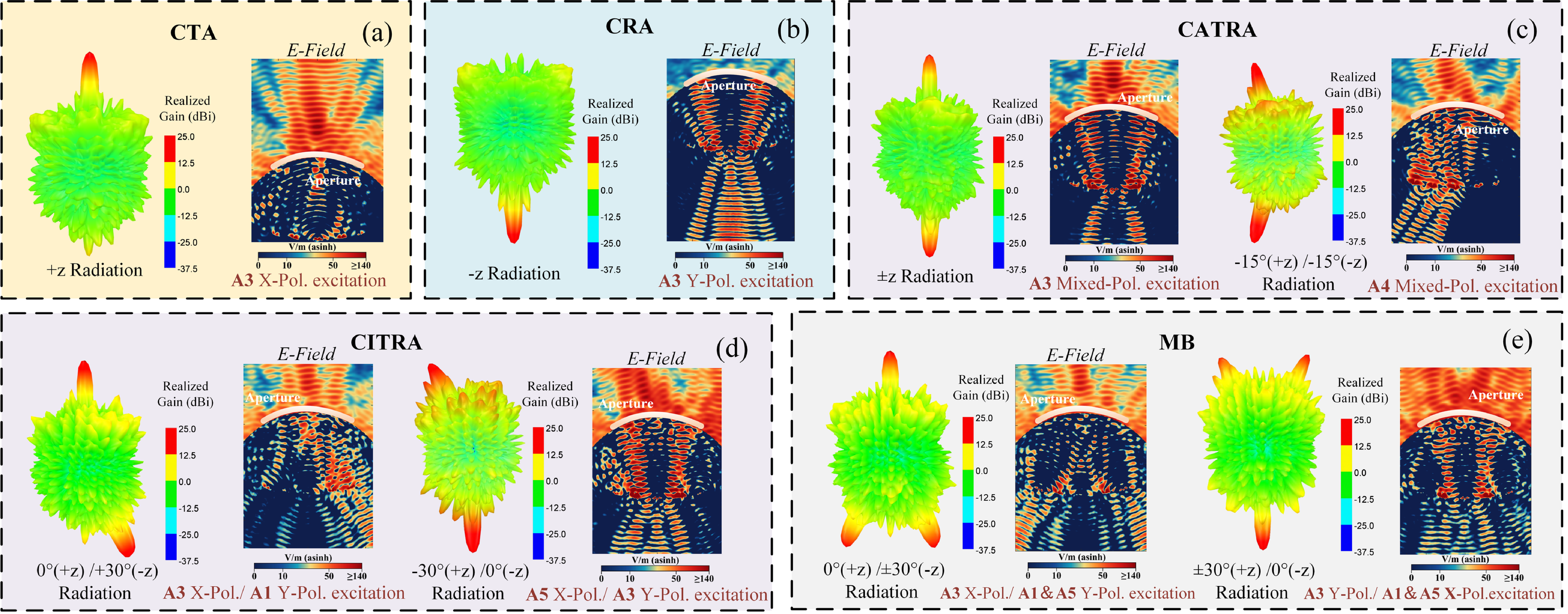}}
		\caption{3-D radiation patterns and E-field distributions for the five operating states. In the E-field maps, the region above the CTRA shows the scattered field, while the region below shows the field difference relative to the case without the CTRA. (a) CTA. (b) CRA. (c) CATRA. (d) CITRA. (e) MB.}
		\label{fig15}
	\end{figure*}
    
\begin{table*}[!t]
\caption{Performance Comparison of Different Operating States}
\centering
\renewcommand{\arraystretch}{1.2}
\setlength{\tabcolsep}{4pt}

\begin{tabular}{|
>{\centering\arraybackslash}m{1.2cm}|
>{\centering\arraybackslash}m{2.6cm}|
>{\centering\arraybackslash}m{1.6cm}|
>{\centering\arraybackslash}m{3.2cm}|
>{\centering\arraybackslash}m{2.6cm}|
>{\centering\arraybackslash}m{2.5cm}|
}
\hline
\textbf{States} &
\textbf{Feed/Excitation Polarization} &
\textbf{No. of Beams} &
\textbf{Beam Switching Capability} &
\textbf{Maximum Switching Angle} &
\textbf{Polarization of Generated Beams} \\
\hline

CTA & x-Pol. & 1 & Forward & $\pm30^{\circ}$ & y-Pol. \\
\hline

CRA & y-Pol. & 1 & Backward & $\pm30^{\circ}$ & y-Pol. \\
\hline

CATRA & Mixed-Pol. & 2 &
Bidirectional (Associative) &
$\pm30^{\circ}/\pm30^{\circ}$ & y-Pol. \\
\hline

CITRA & x-Pol. \& y-Pol. & 2 &
Bidirectional (Independent) &
$\pm30^{\circ}$ and $\pm30^{\circ}$ & y-Pol. \\
\hline

MB & x-Pol. \& y-Pol. & 3 & -- & -- & y-Pol. \\
\hline

\end{tabular}
\end{table*}

As illustrated in Fig. 14, the beam-switching mechanisms of the five operating states are summarized based on the optimized feed arrangement. For the CTA and CRA states in Fig. 14(a) and (b), Port 1 and Port 2 of the selected PSFA are excited to generate the $x$- and $y$-polarized states, respectively. For the CATRA state in Fig. 14(c), both ports are simultaneously excited through a 1-to-2 power divider to produce a $45^\circ$ slant-polarized excitation. As shown in Fig. 14(d), a single SPMT RF switch selects one of A1--A5 for the CTA, CRA, and CATRA states, enabling beam switching among $0^\circ$, $\pm15^\circ$, and $\pm30^\circ$. For the CITRA state in Fig. 14(e), two SPMT RF switches independently select two PSFAs with $x$- and $y$-polarized excitations to control the forward and backward beams, respectively. In addition, three ports from different PSFAs are simultaneously excited with the required polarizations to generate three beams for the MB state. Fig. 15 presents the 3-D radiation patterns and E-field distributions of the five operating states. The CTA and CRA states exhibit forward and backward radiation, respectively. The CATRA and CITRA states show bidirectional beams with associated and independently selected angles, respectively, while the MB state produces three distinct beams. In Fig. 15(a), the field distribution becomes smoother and relatively uniform within the central beam region farther from the aperture along $+z$, despite local variations near the curved surface. In Fig. 15(b), a pronounced field region extends below the aperture along $-z$, consistent with the backward main beam.

	\section{Experiments and Analysis of Conformal Metasurface}
	
	\subsection{Prototyping and Measurement}
	To validate the feasibility of the proposed design, Fig. 16 shows the fabricated prototype with overall dimensions of 300 mm × 376 mm × 232 mm, which was placed in a far-field microwave anechoic chamber for experimental measurement. The antenna support structure was fabricated using 3D-printed resin material, while the metasurface and feed source were assembled and secured with plastic bolts. The prototype fabrication procedure is as follows: first, a 0.1 mm-thick F4BM220 dielectric substrate with a dielectric constant of 2.2 was used, and the metal metasurface pattern was etched on its surface using the OSP (Organic Solderability Preservative) anti-oxidation process. Subsequently, multiple layers of F4B dielectric boards, PDMS dielectric, and PI foam sheets were sequentially stacked and fixed within the cylindrical hollow bracket using plastic bolts and nuts. Similarly, the PSFA antennas were fabricated on Rogers 4003 substrates with a relative permittivity of 3.55 and mounted on the lower bracket using plastic support columns. The overall thickness of the CTRA is 3.8 mm, corresponding to approximately $0.134\lambda$. A PD-2SF-200018000 power divider is employed in the CATRA and CITRA states.

    \begin{figure}[!t]
		\centerline{\includegraphics[width=\columnwidth]{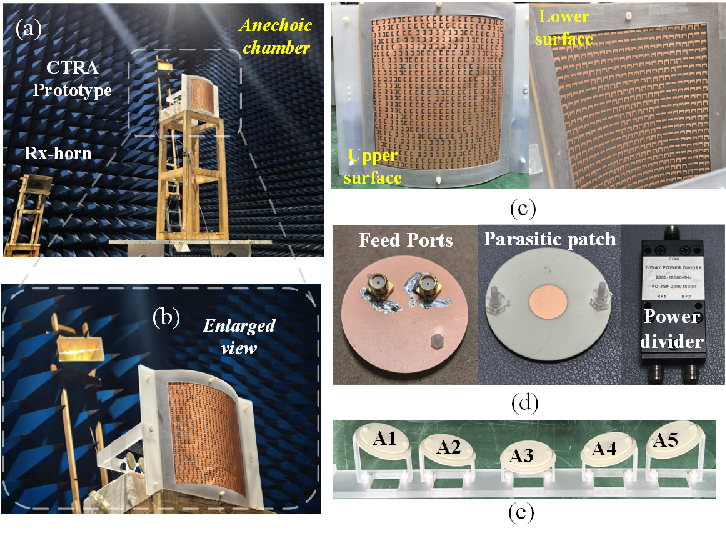}}
		\caption{ (a) Schematic diagram of far-field measurement in a microwave anechoic chamber. (b) Illustration of sample placement. (c) Upper and lower surfaces of CTRAs. (d) PSFA structure and power divider. (e) Arrangement of feed array.}
		\label{fig16}
	\end{figure} 
    
    \begin{figure*}[!t]
		\centerline{\includegraphics[width=0.9\textwidth]{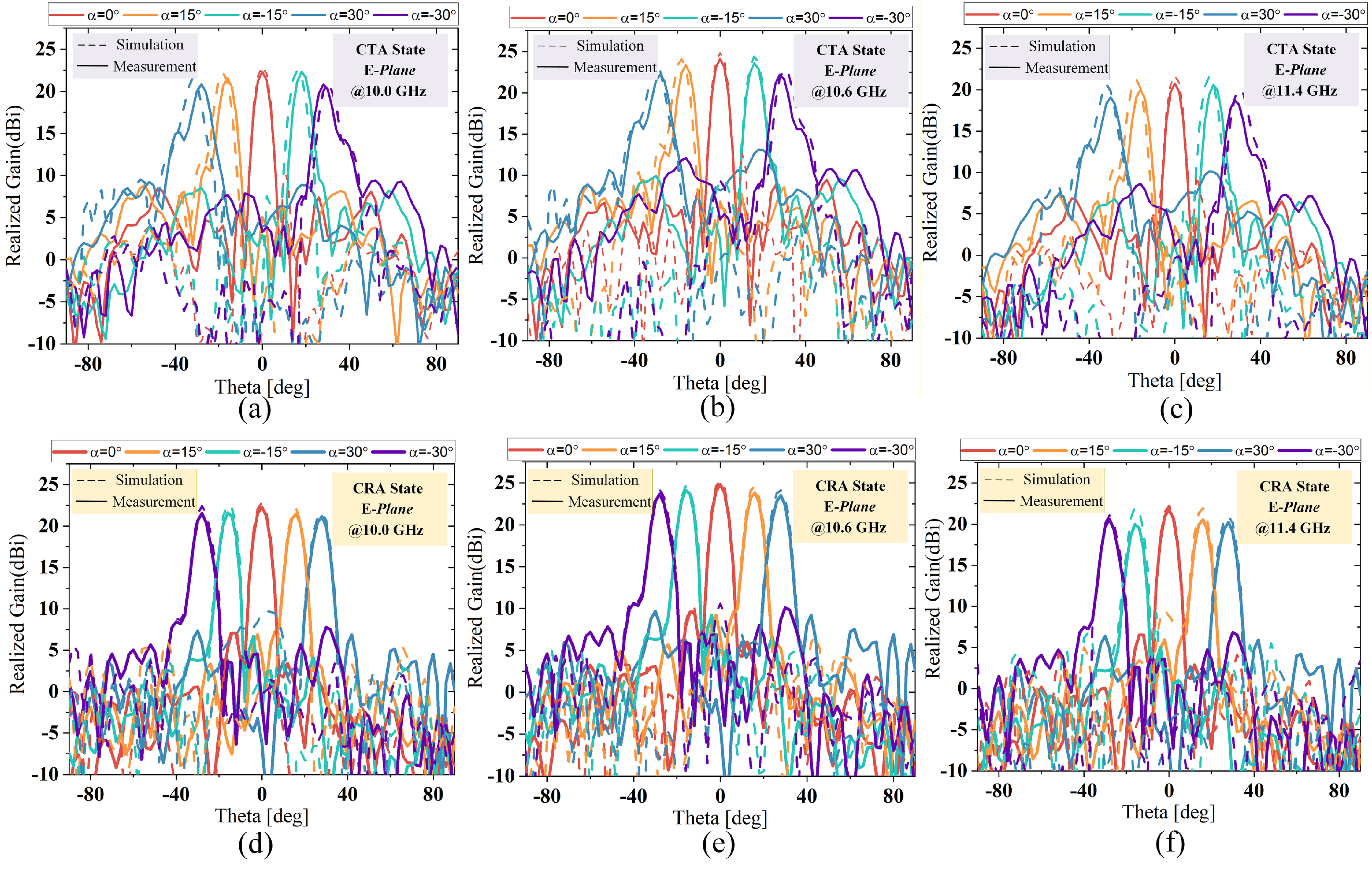}}
		\caption{ Beam switching radiation patterns ranging from 10.0 to 11.4 GHz. (a)-(c) Patterns at CTA state. (d)-(f) Patterns at CRA state.}
		\label{fig17}
	\end{figure*}

    \begin{figure}[!t]
		\centerline{\includegraphics[width=\columnwidth]{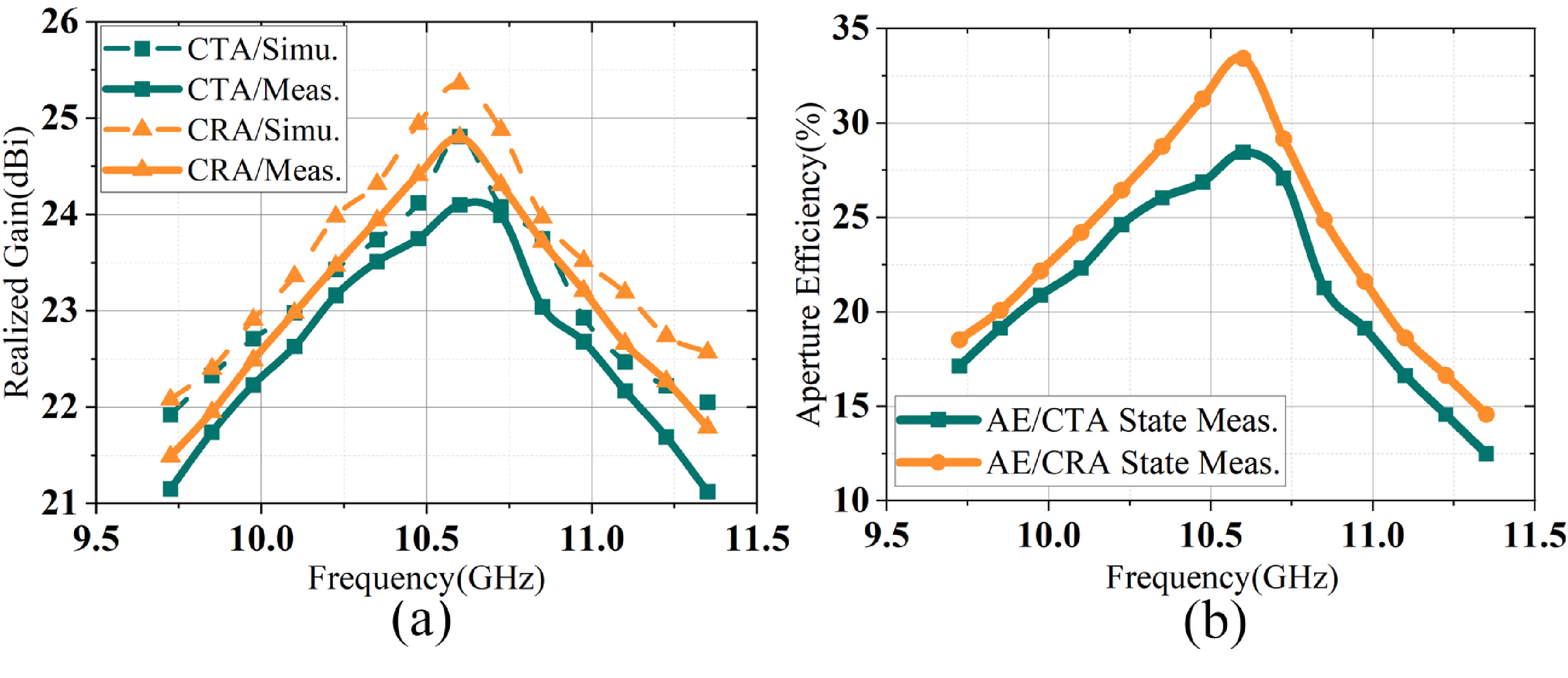}}
		\caption{  (a) Peak gains versus frequency in CTA/CRA states. (b) The measured aperture efficiency in CTA/CRA states.}
		\label{fig18}
	\end{figure} 

    \begin{figure}[!t]
		\centerline{\includegraphics[width=\columnwidth]{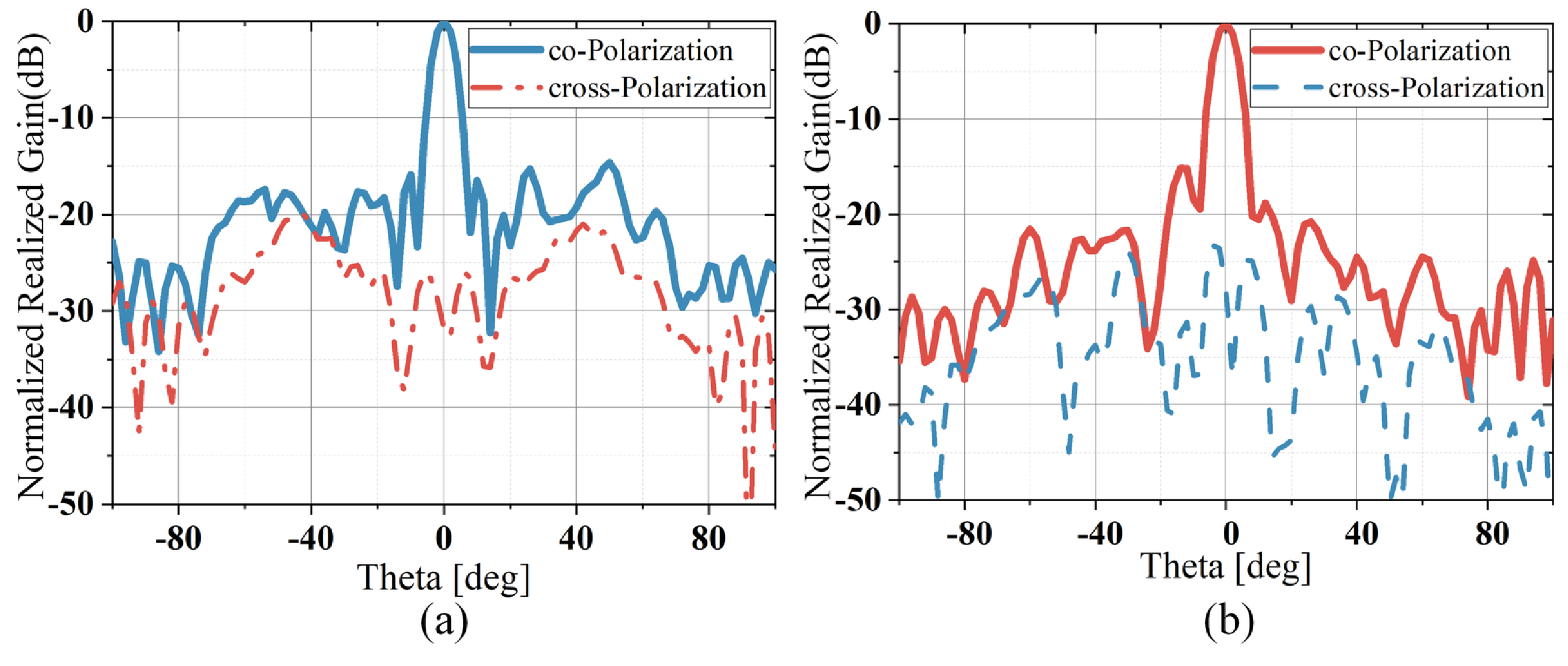}}
		\caption{   Measured co-polarization and cross-polarization radiation patterns: (a) CTA state. (b) CRA state.}
		\label{fig19}
	\end{figure} 

	The PSFA (A1–A5) is arranged in a non-uniform arc configuration, as illustrated in Fig. 16(e). The central feed A3 is designated as the origin of the coordinate system (0, 0, 0) and is oriented perpendicularly to the CTRA surface. The center positions of the remaining feeds are defined as: A1 (0, $-$108 mm, 13 mm), A2 (0, $-$61 mm, 5 mm), A4 (0, 61 mm, 5 mm), and A5 (0, 108 mm, 13 mm). The outer feeds A1 and A5 exhibit an angular offset of approximately 24° relative to the y-axis, while A2 and A4 are offset by approximately 14°. This non-uniform arc arrangement is designed to mitigate phase errors induced by oblique feed incidence and thereby enhance aperture efficiency across various beam switching angles. 
	
	The polarization configurations corresponding to each operating state are summarized in Table I. In the CTA and CRA states, forward and backward beam switching within ±30° is realized, respectively. The CATRA and CITRA states enable bidirectional beam switching in associative and independent modes, both with a maximum switching range of ±30°. In the MB state, three beams are generated simultaneously in several directions. It is noted that all beams produced under the aforementioned states exhibit y-polarized electromagnetic radiation.

      \begin{figure*}[!t]
		\centerline{\includegraphics[width=0.9\textwidth]{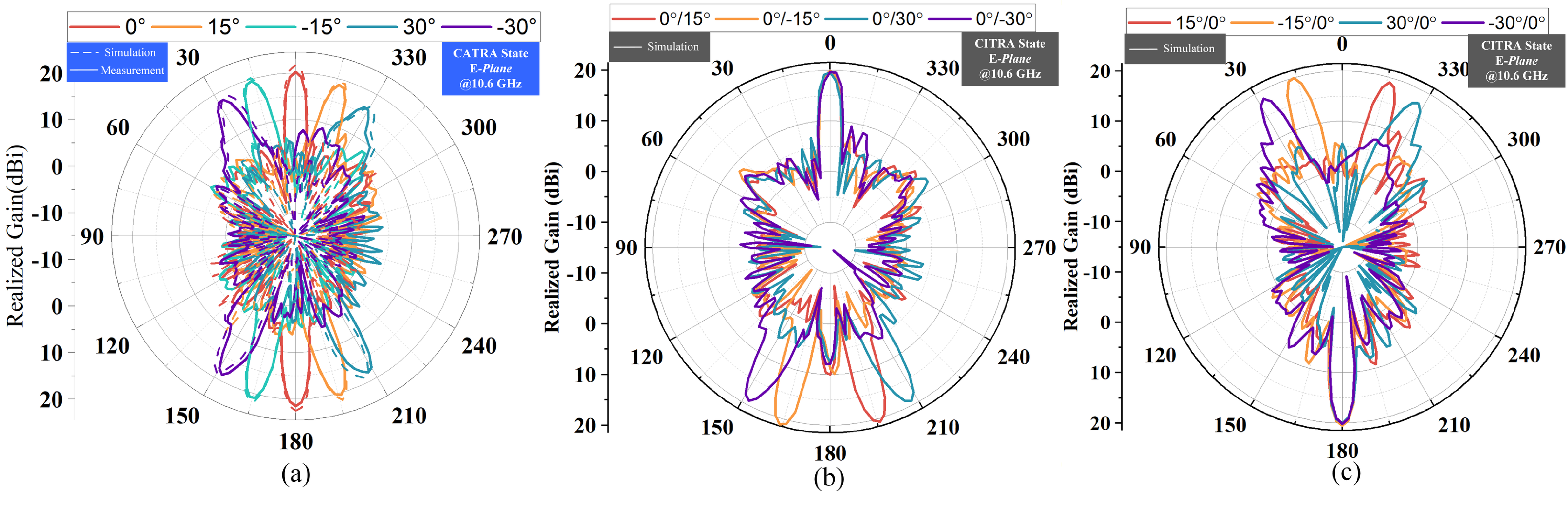}}
		\caption{ Representative bidirectional beam switching radiation patterns: (a) CATRA state. (b) and (c) CITRA state.}
		\label{fig20}
	\end{figure*}

    \begin{figure}[!t]
		\centerline{\includegraphics[width=\columnwidth]{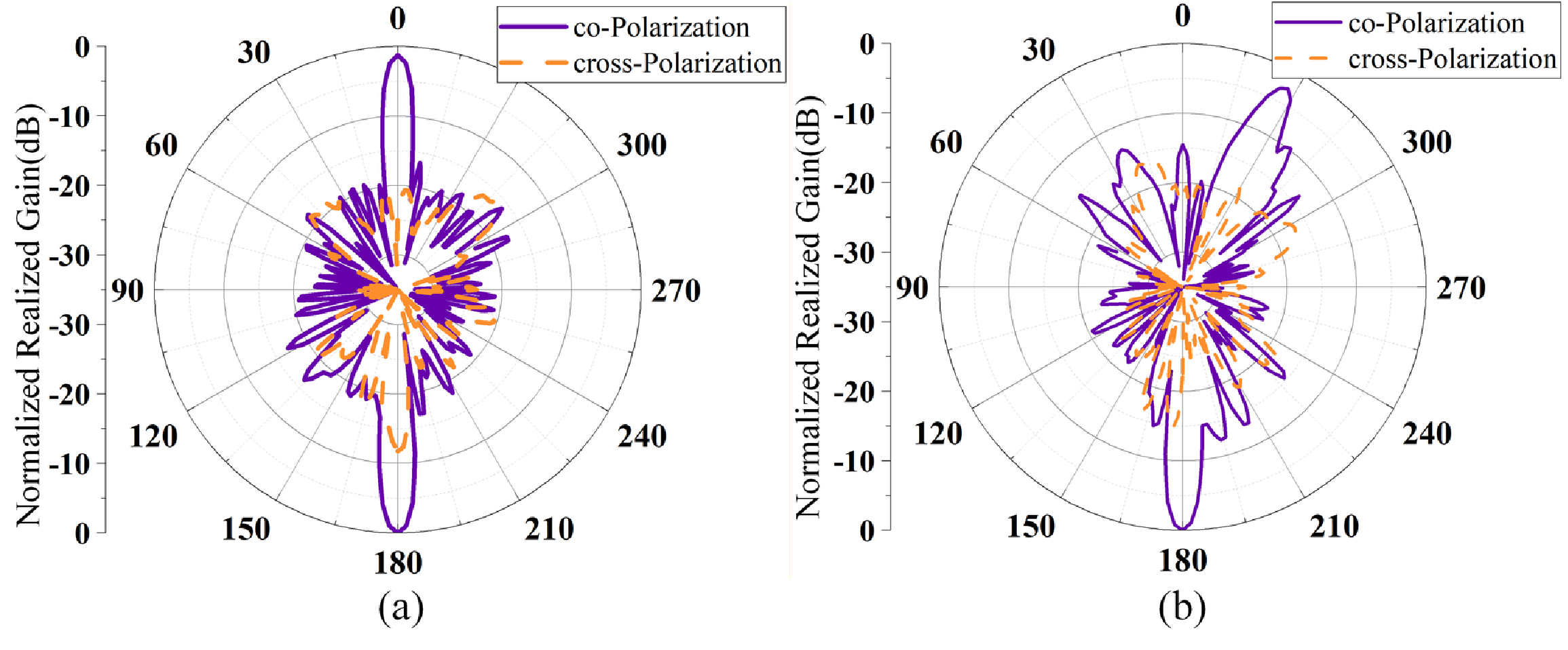}}
		\caption{ Co-polarization and cross-polarization radiation patterns in bidirectional radiation states. (a) CATRA state. (b) CITRA state.}
		\label{fig21}
	\end{figure} 

    \begin{figure}[!t]
		\centerline{\includegraphics[width=\columnwidth]{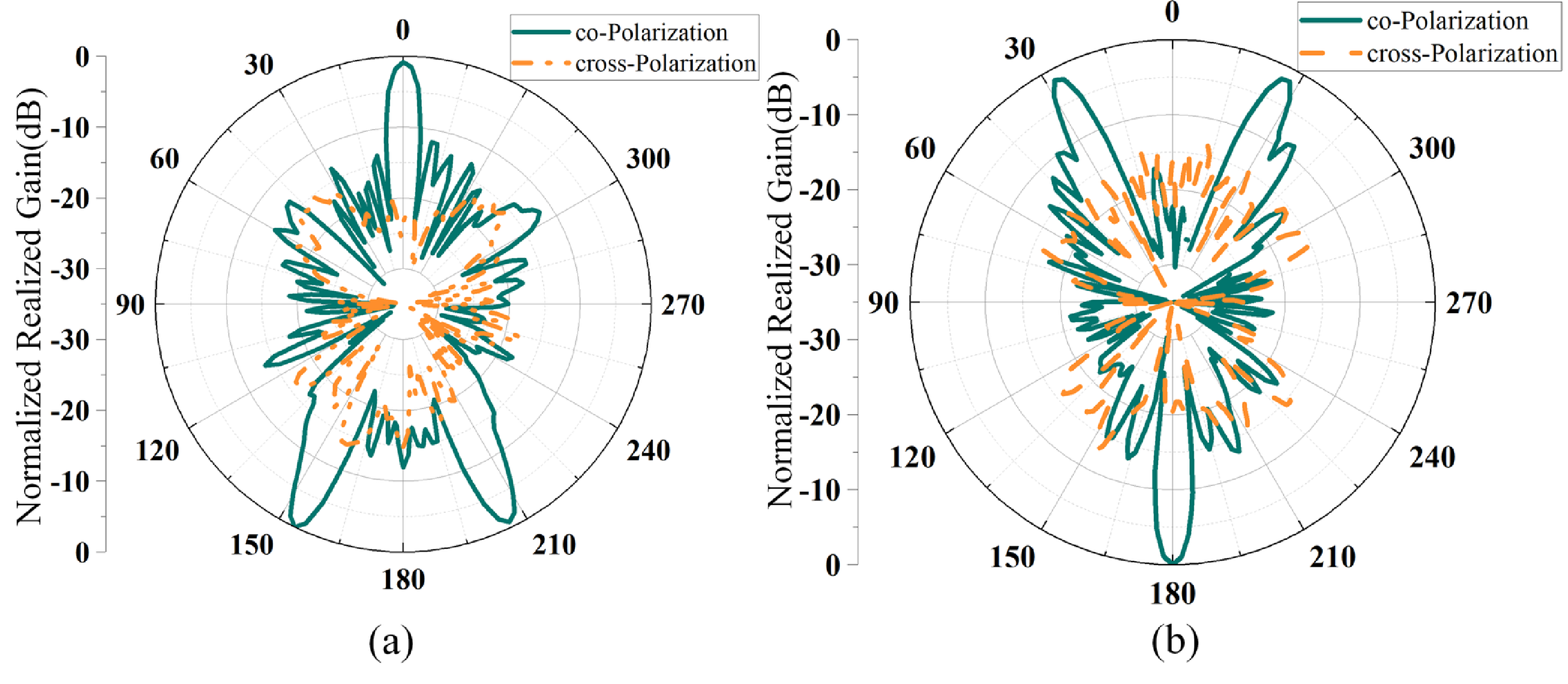}}
		\caption{  Two representative multibeam states: (a) MB1. (b) MB2.}
		\label{fig22}
	\end{figure} 

    \begin{figure}[!t]
		\centerline{\includegraphics[width=\columnwidth]{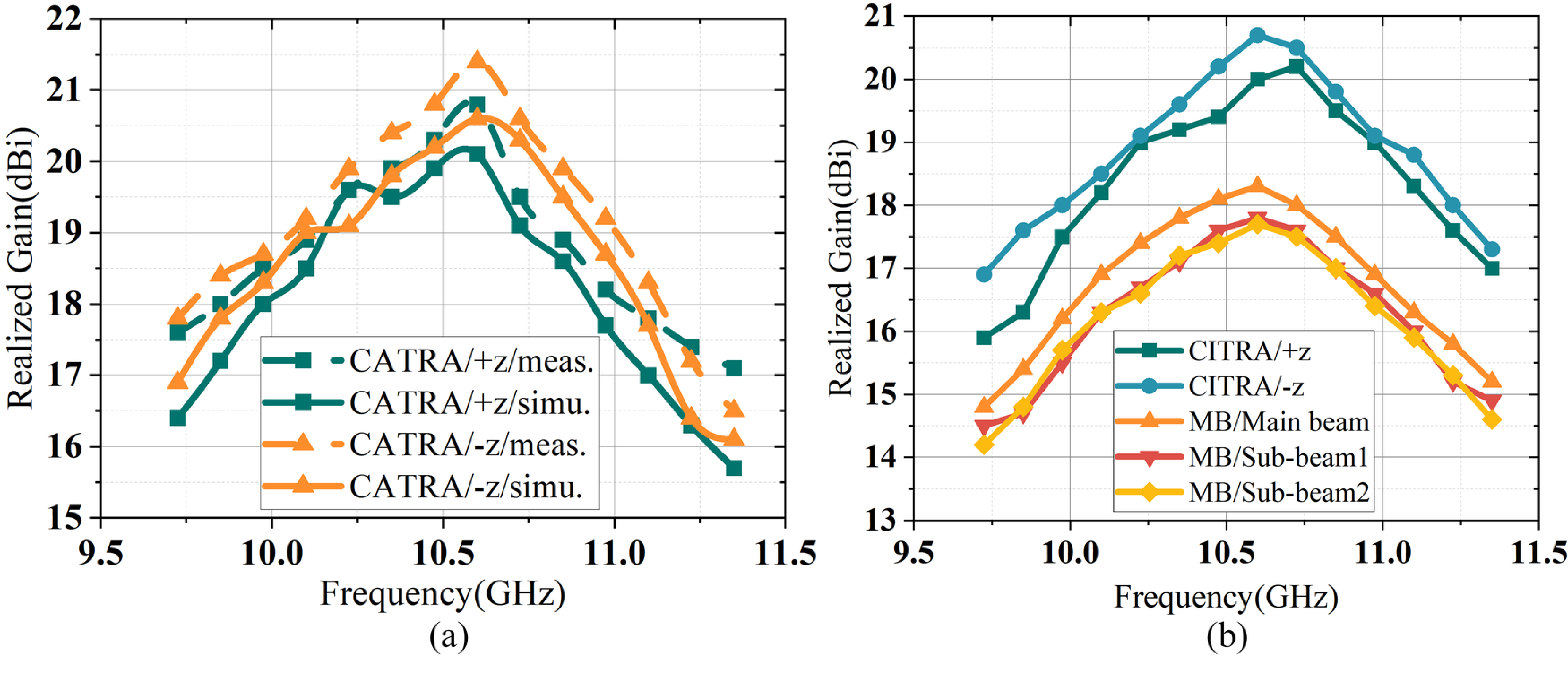}}
		\caption{  (a) Peak gains versus frequency in CATRA state. (b) Simulated peak gains versus frequency in CITRA/MB states.}
		\label{fig23}
	\end{figure} 

	\subsection{CTA and CRA Operation}
	Fig. 17 illustrates the simulated and measured beam-switching performance of the CTA and CRA states at 10.0 GHz, 10.6 GHz, and 11.4 GHz. As shown in Figs. 17(a)–(c), when the CTA state sequentially excites each feed of the PSFA, forward-switching beams of 0°, ±15°, and ±30° are obtained, exhibiting peak gains of 24.1 dBi, 23.6 dBi, and 22.2 dBi at 10.6 GHz, respectively. Similarly, Figs. 17(d)–(f) demonstrate that the CRA state generates reflected beams of 0°, ±15°, and ±30°, with peak gains of 24.8 dBi, 24.1 dBi, and 23.7 dBi at 10.6 GHz, respectively. The maximum gain degradations for the CTA and CRA states are 1.9 dB and 1.1 dB, respectively. Fig. 18 depicts the gain and aperture efficiency fluctuations of the CTA/CRA states near the central operating frequency. For the overall physical aperture of 240 mm × 240 mm on the cylindrical curved surface, the maximum gains realized at 10.6 GHz are 24.1 dBi and 24.8 dBi for the CTA and CRA states, respectively. The corresponding aperture efficiencies are 28.4$\%$ and 33.4$\%$, respectively. The CTA state achieves a 3-dB gain bandwidth of 12.1$\%$ (10.1–11.4 GHz), while the CRA state exhibits a 3-dB gain bandwidth of 13.1$\%$ (10.0–11.4 GHz). The slight performance difference arises mainly from the distinct transmission and reflection efficiencies of the CTRA unit.
	
	Fig. 19 presents the copolarization and cross-polarization radiation patterns in the boresight direction (0°) under the CTA and CRA states. As shown in Fig. 19(a), in the CTA state, the side-lobe level of the E-plane main beam is lower than $-$15.2 dB, while the cross-polarization level does not exceed $-$28.0 dB. In the CRA state, both the side-lobe level and the cross-polarization level at 0° are below $-$15.9 dB and $-$31.8 dB, respectively, as illustrated in Fig. 19(b). In the CRA state, the feed array introduces blockage and scattering, leading to a modest gain reduction and increased sidelobe levels while preserving the main-beam direction. These effects are included in the reported measurements, which were performed with the complete PSFA installed. 

	\subsection{CTRA Operation}
	The CTRA configuration supports both the co-directional (CATRA) and independent-angle (CITRA) operating states. Fig. 20(a) illustrates the radiation patterns corresponding to the CATRA state. By simultaneously exciting the two ports of the same PSFA element to generate hybrid polarization excitation and sequentially activating different feed elements, bidirectional beams at 0$^\circ$, $\pm15^\circ$, and $\pm30^\circ$ are achieved. The peak gains in the forward and backward directions reach 20.1 dBi and 20.6 dBi, respectively. The CITRA state is realized by simultaneously applying different polarization modes to the two feed antennas. It is noteworthy that this design enables a wide range of CITRA operating configurations, of which only several representative cases are presented for demonstration, as shown in Figs. 20(b) and 20(c). In these examples, the simulated radiation results show that the forward 0° and backward 30° beams achieve maximum gains of 20.0 dBi and 19.7 dBi, respectively, while the forward 15° and backward 0° beams achieve maximum gains of 19.4 dBi and 20.7 dBi, respectively.
	Similarly, representative CATRA and CITRA configurations were selected for experimental validation, and their co-polarized and cross-polarized radiation patterns are shown in Fig. 21. For both co-angle and independent-angle bidirectional beam-switching modes, the cross-polarization levels remained below $-$14.7 dB, and the sidelobe levels of the forward and backward beams were effectively suppressed. These results demonstrate that the proposed CTRA enables high-gain directional radiation with flexible bidirectional beam control, supporting both same-angle and independently switched beams in opposite directions.
    \begin{table*}[!t]
\caption{Comparison With Previous Works}
\centering
\footnotesize
\renewcommand{\arraystretch}{1.15}
\setlength{\tabcolsep}{3pt}
\setlength{\arrayrulewidth}{0.6pt}

\newcolumntype{C}[1]{>{\centering\arraybackslash}m{#1}}

\begin{threeparttable}
\begin{tabular}{|
C{0.69cm}|
C{1.28cm}|
C{1.20cm}|
C{0.60cm}|
C{1.50cm}|
C{1.17cm}|
C{2.10cm}|
C{1.90cm}|
C{1.70cm}|
C{1.70cm}|
C{1.70cm}|
}
\hline

\rowcolor[gray]{0.90}
\textbf{Ref.} &
\textbf{\makecell{Frequency\\(GHz)}} &
\textbf{\makecell{Prototype\\Form}} &
\textbf{F/D} &
\textbf{\makecell{Aperture\\Size($\lambda_0$)\tnote{a}}} &
\textbf{\makecell{Gain BW\\(\%)\tnote{b}}} &
\textbf{\makecell{Radiation\\States}} &
\textbf{\makecell{Multiple States\\with Single\\Prototype}} &
\textbf{\makecell{Aperture\\Efficiency(\%)}} &
\textbf{\makecell{Beam\\Switching\\Capability}} &
\textbf{\makecell{Maximum\\Switching\\Angle}}
\\
\hline

{[19]} & 7.3 & TRA &
0.72 &
5.4×5.4 &
12.3/13.7 &
\makecell{$+z$; $-z$} &
Yes(2 States) &
47.2/43.8 &
N.A. &
N.A.
\\
\hline

{[21]} & 30 & TRA &
N.A. &
11.66×11.66 &
N.A. &
\makecell{$\pm z$; MB} &
No &
22.8/22.8 &
N.A. &
N.A.
\\
\hline

{[22]} & 26 & HTA &
0.268 &
8.06×8.06 &
N.A. &
\makecell{$+z$; $-z$; $\pm z$} &
Yes &
33/22.1 &
N.A. &
N.A.
\\
\hline

{[23]} & 5.6/8.8 & FTA &
0.32 &
10.27×10.27 &
12.5/15.9 &
\makecell{$+z$; $-z$} &
Yes(2 States) &
45.9/32.8 &
N.A. &
N.A.
\\
\hline

{[24]} & 9.75 & HTA &
1.067 &
11.7×11.7 &
15.4/15.4 &
\makecell{$+z$; $-z$; $\pm z$} &
Yes(3 States) &
13.3/11.9 &
Bidirectional &
$\pm30^{\circ}/\pm22^{\circ}$
\\
\hline

{[25]} & 9.5 & TRA &
0.583 &
7.6×7.6 &
19.9 &
\makecell{$+z$; $-z$; $\pm z$} &
Yes(3 States) &
19.2/17.3 &
Bidirectional &
$\pm60^{\circ}/\pm45^{\circ}$
\\
\hline

{[31]} & 25 & CTA &
N.A. &
6.55×6.07 &
N.A. &
$+z$ &
N.A. &
25.1 &
Unidirectional &
$\pm15^{\circ}$
\\
\hline

{[32]} & 15 & CTA &
0.7 &
10×6 &
6.3 &
$+z$ &
N.A. &
15 &
Unidirectional &
$\pm50^{\circ}$
\\
\hline

{[38]} & 28 & CTA/CRA &
1.076 &
9.24×8.68 &
9.8/4.6 &
\makecell{$+z$; $-z$; $\pm z$} &
No &
36/35 &
N.A. &
N.A.
\\
\hline

{[41]} & 5.35 & TRA &
1 &
7.13×7.13 &
32/32 &
$\pm z$ &
N.A. &
8.2/5.4 &
Bidirectional &
$\pm60^{\circ}/\pm60^{\circ}$
\\
\hline

\textbf{\makecell{This\\Work}} &
\textbf{10.6} &
\textbf{CTRA} &
\textbf{0.917} &
\textbf{8.48×8.48} &
\textbf{12.1/13.1} &
\makecell{
    \textbf{$+z$; $-z$; $\pm z(A\tnote{*})$;}\\
    \textbf{$\pm z(I\tnote{**})$; MB}
} &
\textbf{Yes(5 States)} &
\textbf{28.4/33.4} &
\makecell{
    \textbf{Independent}\\
    \textbf{Bidirectional}
} &
\textbf{$\pm30^{\circ}/\pm30^{\circ}$}
\\
\hline

\end{tabular}

\begin{tablenotes}
\item[*] Associate angle beam switching;
\item[**] Independent angle beam switching.

\item[a]

Aperture size is expressed as
$(L_x/\lambda_0)\times(L_y/\lambda_0)$
for a rectangular aperture, or
$D/\lambda_0$ (diameter) for a circular aperture,
where $\lambda_0$ is the free-space wavelength
at the listed frequency.

\item[b]

Gain BW: Gain BW denotes the 3-dB gain bandwidth.

\end{tablenotes}
\end{threeparttable}
\end{table*}

   \subsection{MB State Operation}
  The proposed design supports simultaneous excitation of multiple feed elements for multibeam radiation. As shown in Figs.~22(a) and (b), the MB1 and MB2 states each generate three beams. In MB1, A1 and A5 are excited with $y$-polarization and A3 with $x$-polarization, producing beams along $+z$ and at $\pm30^\circ$ relative to $-z$, with peak gains of 17.2, 17.8, and 17.7~dBi, respectively. In MB2, A1 and A5 are excited with $x$-polarization and A3 with $y$-polarization, producing beams along $-z$ and at $\pm30^\circ$ relative to $+z$, with peak gains of 18.3, 16.8, and 16.6~dBi, respectively. The polarization isolation between the transmission and reflection channels supports simultaneous bidirectional multibeam operation. Since the beam directions are feed-selected and discrete, reduced-gain crossover regions may occur between adjacent beams. Fig. 23 shows that representative CATRA, CITRA, and MB beams achieve 3-dB gain bandwidths of approximately 11.6\%, 12.9\%, and 13.9\%, respectively.

	\subsection{Discussion and Comparison}
	To further demonstrate the innovation of the proposed CTRA, Table II summarizes a detailed comparison with existing works. The proposed CTRA also achieves comparable gain bandwidths (12.1\%/13.1\%) with an $F/D$ of 0.917 and a normalized aperture width of 8.48$\lambda_0$. Benefiting from the proposed complementary-symmetric transmit–reflect integrated element, the presented design realizes bidirectional multibeam switching in a conformal metasurface aperture, while offering flexible beam control and multiple operating modes. These features significantly enhance its adaptability to complex and dynamic emergency communication scenarios. To the best of the authors’ knowledge, this approach has not been previously reported. Unlike previous studies, this approach eliminates the need for numerous PIN diodes or bulky design configurations, thereby simplifying the system control mechanism. Although multiple feeds are used, beam switching is realized through an RF switching network rather than independent RF chains, resulting in only limited additional complexity. Compared with the CTA/CRA architectures reported in [38], the proposed design integrates both transmission and reflection functions within a single prototype, allowing wide-angle switching among multiple radiation states. Furthermore, unlike the bidirectional HTA/TRA architectures with beam-switching capabilities reported in [24] and [25], the proposed CTRA achieves a simpler implementation while maintaining higher aperture efficiency. More importantly, it integrates co-angle bidirectional switching, independent-angle beam control, and multibeam radiation within a single conformal aperture, enabling flexible and adaptive beam management. Such multifunctional capability provides great potential for emergency communication applications requiring rapid deployment and reliable bidirectional links under complex propagation environments. 
	
	\section{Conclusion}
	This paper presents an X-band conformal transmit–reflect array (CTRA) capable of multi-state bidirectional beam switching. By integrating a transmission–reflection independently controlled metasurface with a polarization-switchable feed array (PSFA), independent beam switching is achieved in the (+z) and (-z) directions. Moreover, simultaneous multi-polarization excitation of the PSFA enables bidirectional beam switching at identical or independently controlled angles, as well as bidirectional multibeam radiation. The measured maximum gains in the forward and backward directions are 24.1 and 24.8 dBi, corresponding to aperture efficiencies of 28.4$\%$ and 33.4$\%$, respectively. A simplified prototype is developed to demonstrate five distinct operating states, including unidirectional, bidirectional, and multibeam modes, thereby validating the multifunctional and flexible switching capability of the proposed design through combined experimental and simulated results. Benefiting from its lightweight structure, conformal integration capability, high gain, and efficient radiation performance, the proposed CTRA exhibits significant potential for deployment in emergency communication networks and integrated sensing and communication (ISAC) systems.

\begin{IEEEbiography}[{\includegraphics[width=1in,height=1.25in,clip,keepaspectratio]{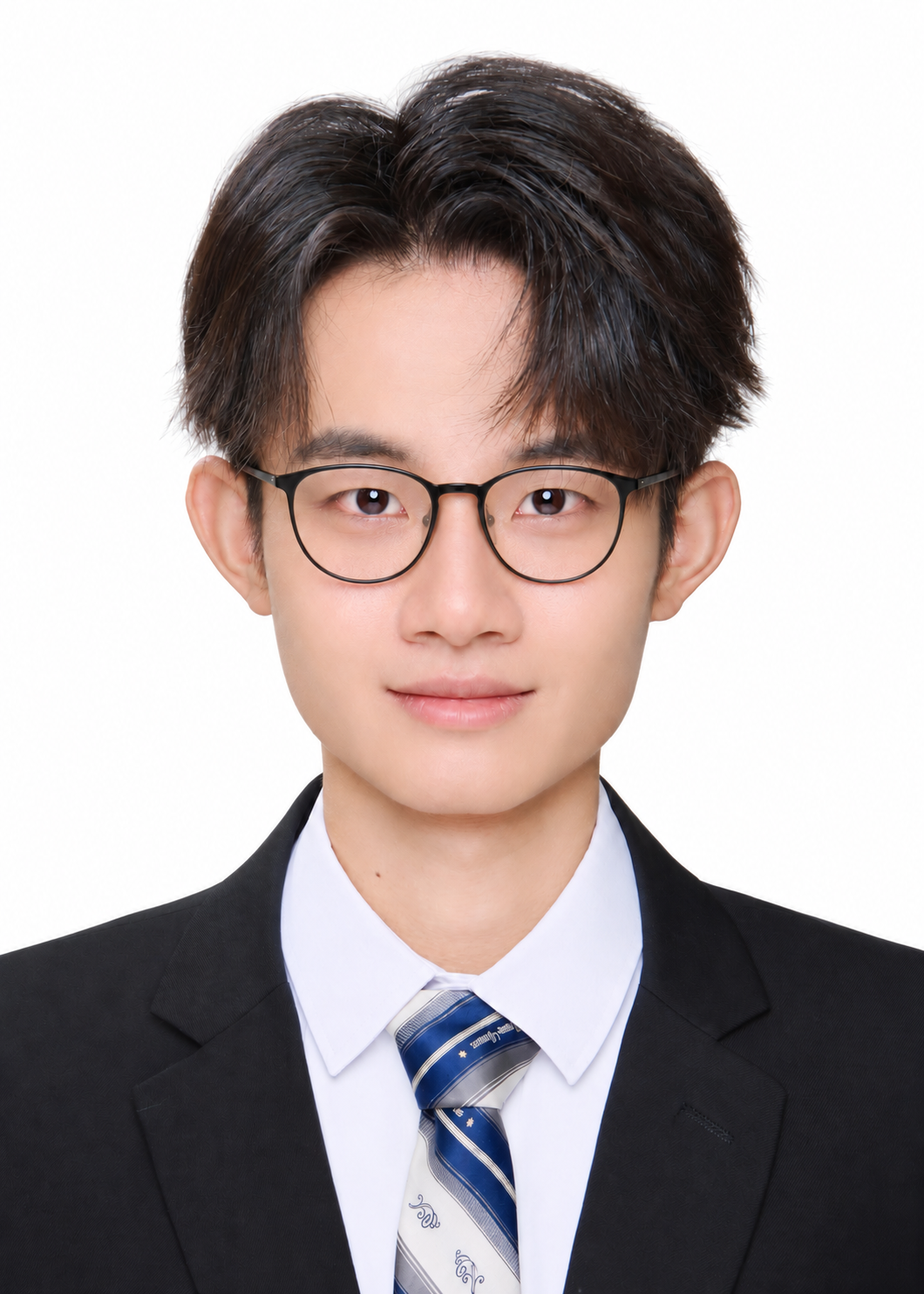}}]{Yifei Liu} (Student Member, IEEE) was born in Luohe, Henan, China. He received the B.S. degree in communication engineering from Zhengzhou University, Zhengzhou, China, in 2022. He is currently pursuing the Ph.D degree in information and telecommunication engineering with Xidian University, Xi’an, China. 

His current research interests involve the transmitarray/reflectarray and bidirectional antennas.
\end{IEEEbiography}

\begin{IEEEbiography}[{\includegraphics[width=1in,height=1.25in,clip,keepaspectratio]{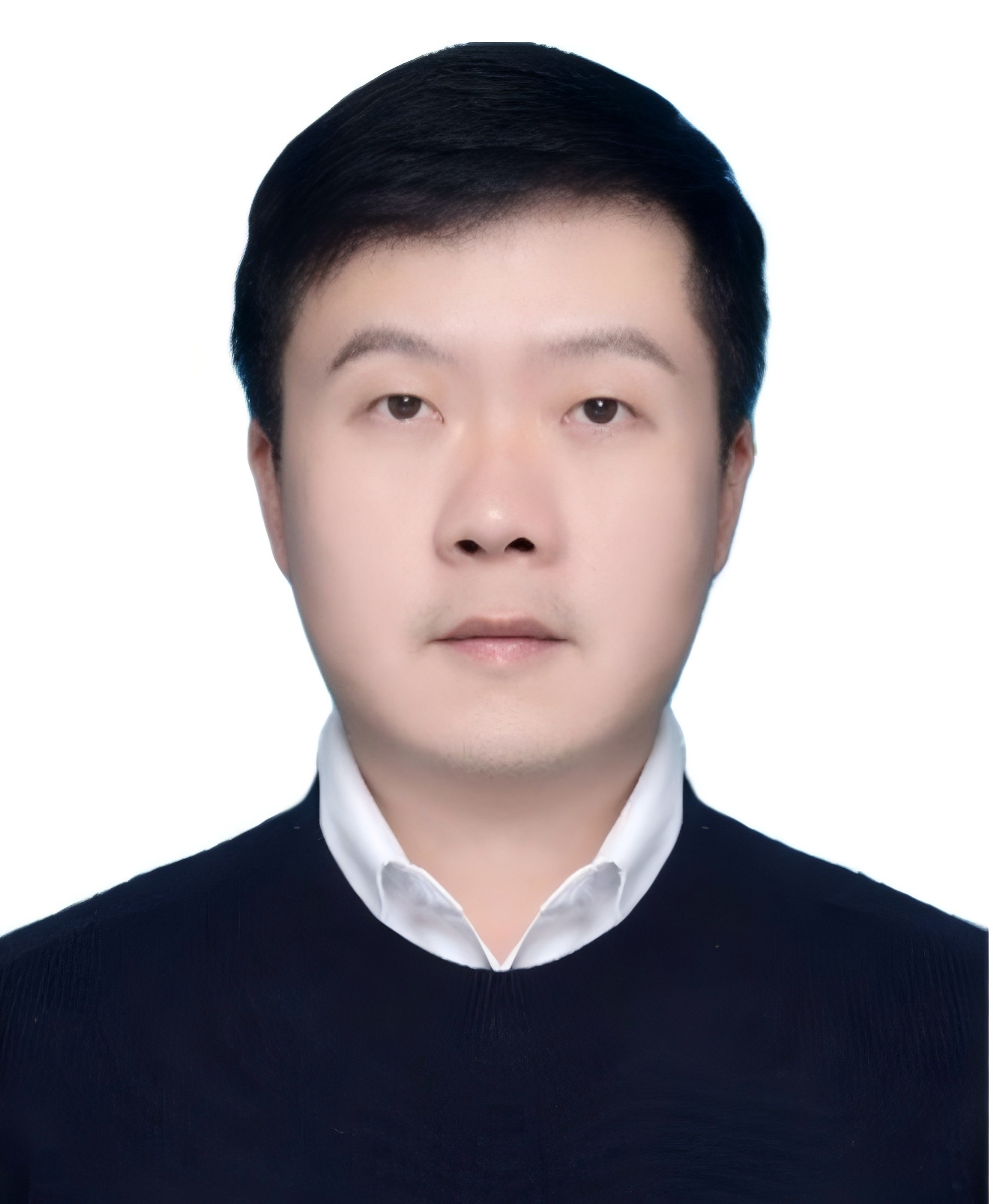}}]{Fan Qin} (Member, IEEE) received the B.S. degree in electronic information engineering and the Ph.D. degree in electromagnetic wave and microwave technology from the Northwestern Polytechnical University, Xi’an, China, in 2010 and 2016, respectively. 

He was a Visiting Scholar with the University of Kent, Canterbury, U.K., from 2013 to 2015. He joined the School of Telecommunications Engineering, Xidian University, Xi’an, in 2016, where he is currently an Associate Professor. His research interests involve the metamaterial antennas, millimeter wave antennas, vortex wave antennas, and metasurface-assisted wireless communication.
\end{IEEEbiography}

\begin{IEEEbiography}[{\includegraphics[width=1in,height=1.25in,clip,keepaspectratio]{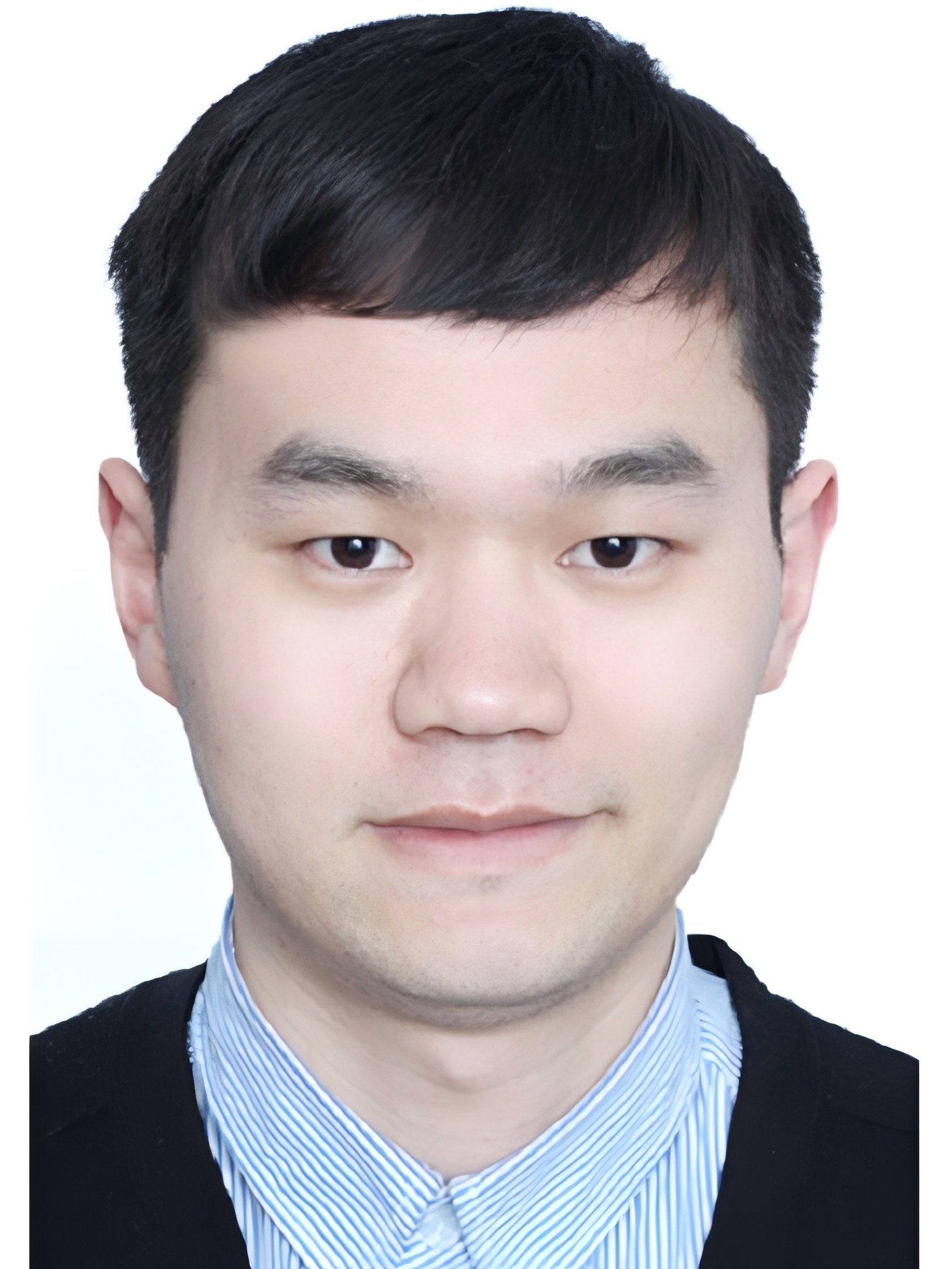}}]{Chao Gu} (Member, IEEE) received the B.Sc. and M.Sc. degrees from Xidian University, Xi’an, China, in 2009 and 2012, respectively, and the Ph.D. degree from the University of Kent, Canterbury, U.K., in 2017. 

From 2017 to 2018, he was a Research Associate with the University of Kent, where he led the development of additive manufacturing techniques for radio frequency components and antennas. Since August 2018, he has been a Senior Research Engineer with the Centre for Wireless Innovation, School of Electronics, Electrical Engineering and Computer Science, Queen’s University Belfast, Belfast, U.K., actively participating in and leading extensive research and development tasks for various wireless applications, including space-based instrumentation and imaging, wireless power transfer, and subterahertz wireless solutions. His research interests include reconfigurable antennas, metasurface antennas, and phased arrays.
\end{IEEEbiography}

\begin{IEEEbiography}[{\includegraphics[width=1in,height=1.25in,clip,keepaspectratio]{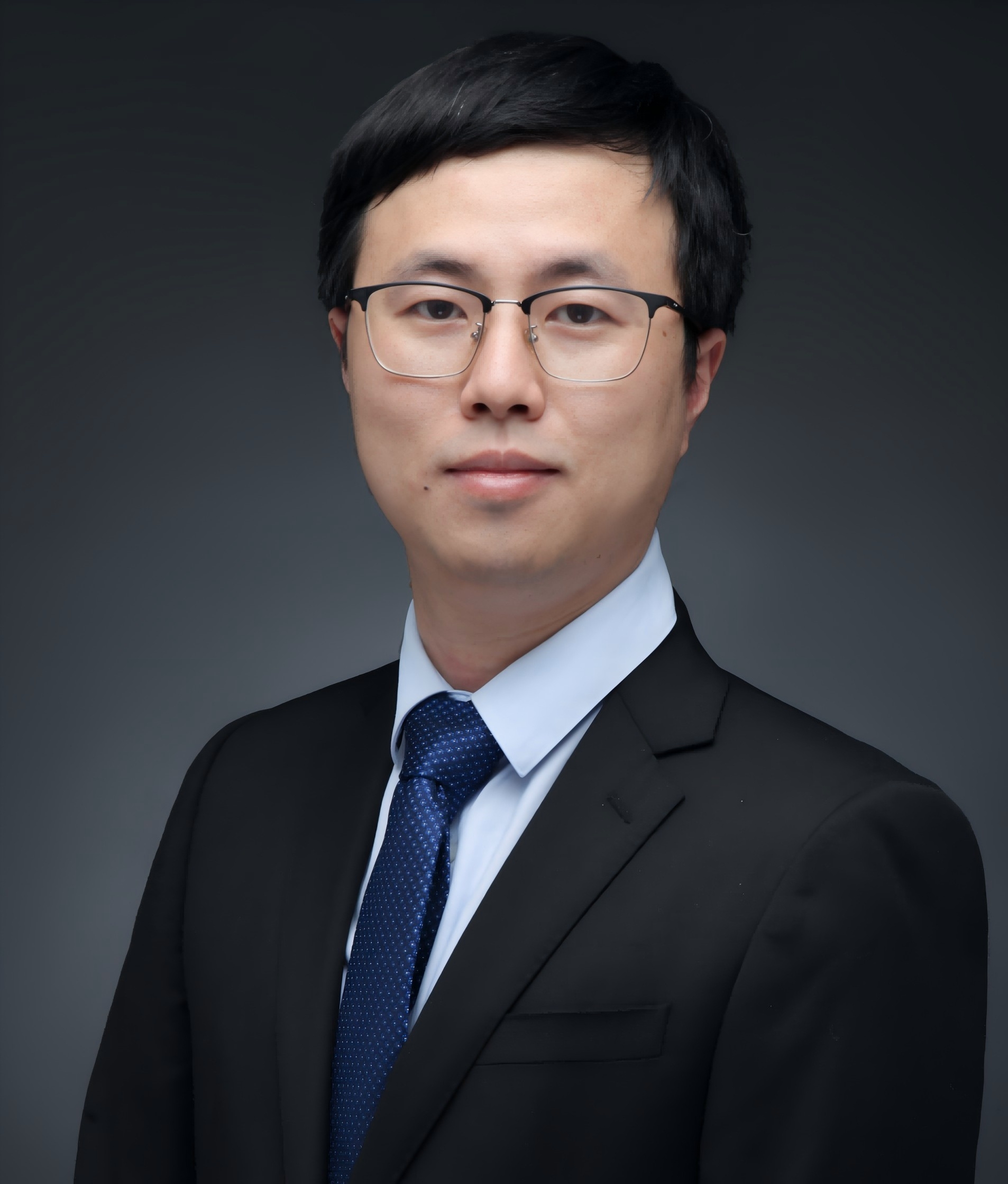}}]{Wenchi Cheng} (Senior Member, IEEE) received B.S. and Ph.D. degrees in telecommunication engineering from Xidian University, Xi’an, China, in 2008 and 2013, respectively.

He was a Visiting Scholar with the Department of Electrical and Computer Engineering, Texas A\&M University, College Station, TX, USA, from 2010 to 2011. He is currently a Full Professor with the Xidian University. He has authored more than 200 international journal and conference papers in IEEE \textsc{Journal on Selected Areas in Communications}, \textit{IEEE} \textit{Magazines}, IEEE \textsc{Transactions}, IEEE INFOCOM, GLOBECOM, and ICC. His current research interests include 6G wireless networks, electromagnetic-based wireless communications, and emergency wireless communications and systems. 

He received the IEEE ComSoc Distinguished Lecturer, the IEEE ComSoc Asia-Pacific Outstanding Young Researcher Award, and four IEEE journal/ conference best papers. He has served or serving as the IEEE GLOBECOM 2026 Tutorial Co-Chair, the Communication Theory Symposium Co-Chair for IEEE ICC 2026, the Wireless Communications Symposium Co-Chair for IEEE ICC 2022 and IEEE GLOBECOM 2020, the Publicity Chair for IEEE ICC 2019. He has served or serving as the ComSoc Representative for the IEEE Public Safety Technology Initiative and an Editor for IEEE \textsc{Transactions On Wireless Communications}.
\end{IEEEbiography}

\begin{IEEEbiography}[{\includegraphics[width=1in,height=1.25in,clip,keepaspectratio]{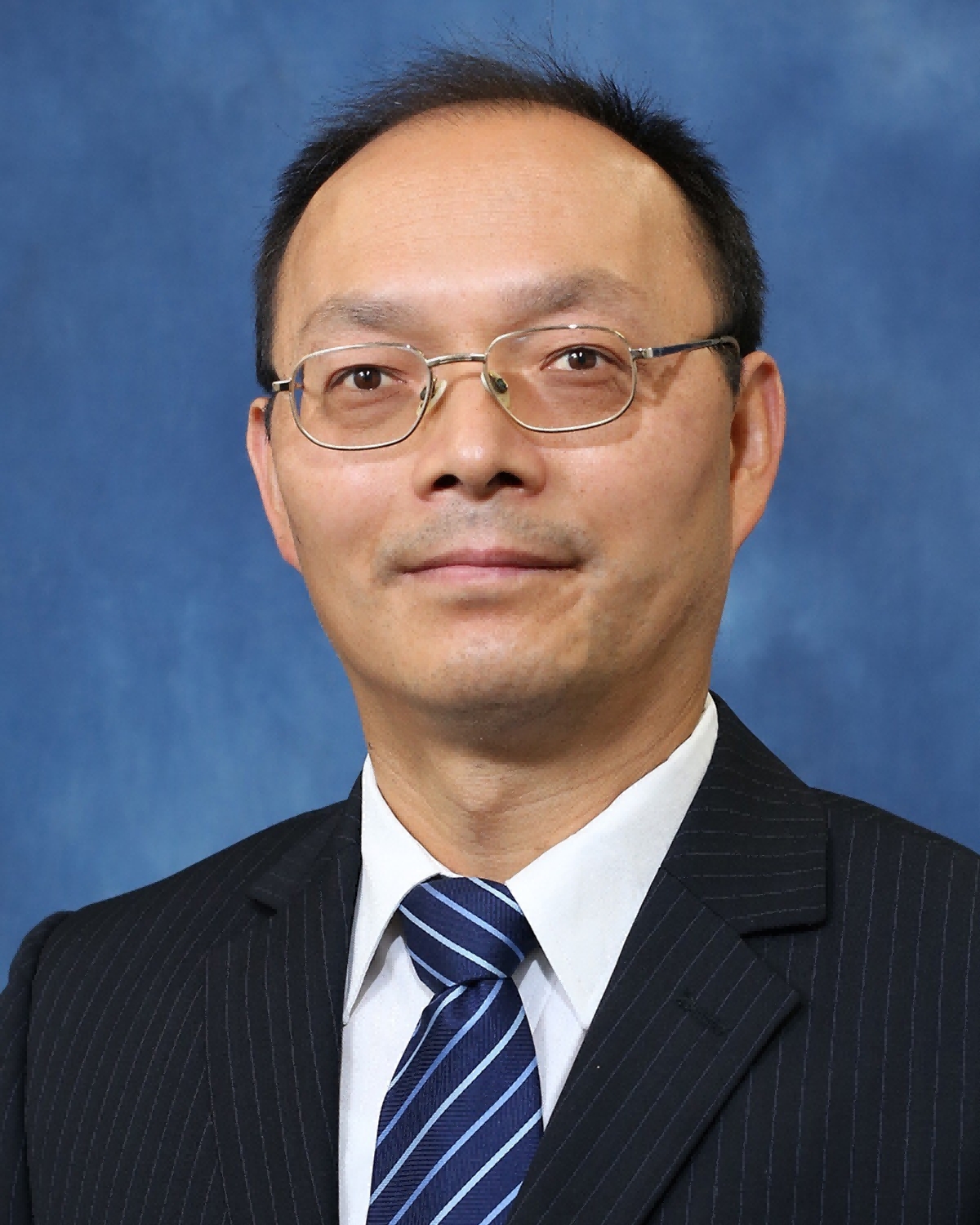}}]{Steven Gao} (Fellow, IEEE) received the Ph.D. degree from Shanghai University, Shanghai, China, in 1999. 

He was a Professor and the Chair of RF/microwave engineering with the University of Kent, Canterbury, U.K., from 2013 to 2022, where he has been an Honorary Professor since 2022. He is currently a Professor with the Department of Electronic Engineering, The Chinese University of Hong Kong, Hong Kong, where he is also the Director of the Center for Intelligent Electromagnetic Systems. He has co-authored/co-edited three books [\textit{Space Antenna Handbook} (Wiley, 2012), \textit{Circularly Polarized Antennas} (IEEE \& Wiley, 2014), and \textit{Low-Cost Smart Antennas} (Wiley, 2019)], and more than 500 articles and holds 20 patents. His research covers smart antennas, phased arrays, MIMO, reconfigurable antennas, broadband/multiband antennas, satellite antennas, and wireless systems (mobile and satellite communications, synthetic aperture radars, and IoT). 

Dr. Gao was U.K.’s Representative in the European Association on Antennas and Propagation (EurAAP) from 2021 to 2022, the General Chair or the General Co-Chair for international conferences (LAPC 2013 and UCMMT 2021), and an Invited/Keynote Speaker at many conferences. He is the TPC Chair for ISAPE 2024. He is the Editor-in-Chief of IEEE \textsc{Antennas And Wireless Propagation Letters}. He was a Distinguished Lecturer of the IEEE Antennas and Propagation Society from 2014 to 2016 and an Associate Editor of several international journals, such as IEEE \textsc{Transactions On Antennas And Propagation} (TAP), \textit{Radio Science}, \textit{Electronics Letters}, and \textit{IET Circuits}, \textit{Devices and Systems}. He served as the Lead Guest Editor for \textsc{Proceedings of the} IEEE for a Special Issue on “Small Satellites” in 2018 and IEEE TAP for Special Issues on “Low-Cost Wide-Angle Beam Scanning Antennas” in 2022 and “Antennas for Satellite Communication” in 2015, and a Guest Editor for \textit{IET Circuits}, \textit{Devices and Systems} for a Special Issue on “Photonic and RF Communications Systems” in 2014.
\end{IEEEbiography}

\end{document}